\documentclass[twocolumn]{aastex631}
\usepackage{comment}

\usepackage{amsmath,amssymb}
\shortauthors{M. De Furio et al.}
\shorttitle{M-type Stellar Binaries in the ONC}
\graphicspath{{./}{figures/}}
\begin{document}

\title{Demographics of the M-star Multiple Population in the Orion Nebula Cluster}

\correspondingauthor{Matthew De Furio}
\email{defurio@umich.edu}

\author[0000-0003-1863-4960]{Matthew De Furio}
\affiliation{Department of Astronomy, University of Michigan, Ann Arbor, MI 48109, USA}

\author{Christopher Liu}
\affiliation{Department of Astronomy, University of Michigan, Ann Arbor, MI 48109, USA}

\author{Michael R. Meyer}
\affiliation{Department of Astronomy, University of Michigan, Ann Arbor, MI 48109, USA}

\author{Megan Reiter}
\affiliation{Rice University, Houston, TX}

\author{Adam Kraus}
\affiliation{Department of Astronomy, University of Texas at Austin, Austin, TX 78712, USA}

\author{Trent Dupuy}
\affiliation{Institute for Astronomy, University of Edinburgh, Royal Observatory, Blackford Hill, Edinburgh, EH9 3HJ, UK}

\author{John Monnier}
\affiliation{Department of Astronomy, University of Michigan, Ann Arbor, MI 48109, USA}

%\author{Alexandra Greenbaum}
%\affiliation{Department of Astronomy, University of Michigan, Ann Arbor, MI}

%\author{Megan Reiter}
%\affiliation{Rice University, Houston, TX}

%\author{John Monnier}
%\affiliation{Department of Astronomy, University of Michigan, Ann Arbor, MI}

%\author{Adam Kraus}
%\affiliation{Department of Astronomy, University of Texas - Austin, Austin, TX}

%\author{Trent Dupuy}
%\affiliation{University of Edinburgh, Edinburgh, United Kingdom}



\begin{abstract}

We present updated results constraining multiplicity demographics for the stellar population of the Orion Nebula Cluster (ONC, a high-mass, high-density star-forming region), across primary masses 0.08-0.7M$_{\odot}$. Our study utilizes archival Hubble Space Telescope data obtained with the Advanced Camera for Surveys using multiple filters (GO-10246). Previous multiplicity surveys in low-mass, low-density associations like Taurus identify an excess of companions to low-mass stars roughly twice that of the Galactic field and find the mass ratio distribution consistent with the field. Previously, we found the companion frequency to low-mass stars in the ONC is consistent with the Galactic field over mass ratios=0.6-1.0 and projected separations=30-160au, without placing constraints on the mass ratio distribution. In this study, we investigate the companion population of the ONC with a double point-spread function (PSF) fitting algorithm sensitive to separations larger than 10au (0.025”) using empirical PSF models. We identified 44 companions (14 new), and with a Bayesian analysis, estimate the companion frequency to low-mass stars in the ONC =0.13$^{+0.05}_{-0.03}$ and the power law fit index to the mass ratio distribution =2.08$^{+1.03}_{-0.85}$ over all mass ratios and projected separations of 10-200au. We find the companion frequency in the ONC is consistent with the Galactic field population, likely from high transient stellar density states, and a probability of 0.002 that it is consistent with that of Taurus. We also find the ONC mass ratio distribution is consistent with the field and Taurus, potentially indicative of its primordial nature, a direct outcome of the star formation process.

\end{abstract}



\section{Introduction} \label{sec:intro}

Stellar multiple systems are a frequent outcome of star formation, and most stars form in clusters or associations with initial density much higher than the disk of the Milky Way.  They are thought to be created through two dominant mechanisms, turbulent fragmentation \citep{Goodwin2004, Offner2010} and disk fragmentation \citep{Adams1989, Bonnell1994, Kratter2008}.  Their properties (e.g. separation, mass ratio (m$_{companion}$/m$_{primary}$), and eccentricity) can be altered by various processes.  The mass ratio can increase due to an inwardly migrating companion preferentially accreting material from the circumstellar disk \citep{Mazeh1992ApJ...401..265M, Kroupa1995MNRAS.277.1507K, Bate1997MNRAS.285...33B, Bate2000MNRAS.314...33B}.  Wide companions can migrate to closer separations through interactions with infalling gas from the protostellar cloud or circumstellar disk \citep{Bate2002MNRAS.336..705B, Bate2003MNRAS.339..577B, Offner2010, Bate2012MNRAS.419.3115B}.  Other processes, like dynamical interactions between a multiple system and other cluster members, can even cause the complete dissolution of the multiple system \citep{Kroupa2001}.

%%% MOVE TO DISCUSSION SECTION %%%
%They also find mass ratio distributions well-fitted with a power law whose exponent ($\beta$) appears to increase with decreasing primary mass: $\sim$ flat ($\beta$ = 0) for solar-type primaries (0.7 - 1.3 M$_{\odot}$) \citep{Raghavan2010} and stellar M-type primaries ($>$ 0.3 M$_{\odot}$) \citep{ReggianiMeyer2011, ReggianiMeyer2013}, 3.6 for very-low mass star and brown dwarf primaries (0.04 - 0.09 M$_{\odot}$) \citep{Reid2006}, and 6.1 for T5-Y0 brown dwarf primaries (0.017 - 0.061 M$_{\odot}$) \citep{Fontanive2018}.
%%% MOVE TO DISCUSSION SECTION %%%

Previous multiplicity surveys in the Galactic field have characterized the companion population as a function of primary mass \citep{DucheneKraus2013, Offner2022arXiv220310066O}.  They find a separation distribution well-fitted with a log-normal where the mean separation appears to increase with primary mass: 6 au for very-low mass star and brown dwarf primaries \citep{Reid2006}, 20 au for stellar M-type primaries \citep{Janson2012, Winters2019AJ....157..216W}, and 50 au for solar-type primaries \citep{Raghavan2010}.  

Multiplicity surveys in young, nearby associations find an excess of companions to low-mass primary stars relative to the Galactic field \citep{Ghez1993,Leinert1993,ReipurthZinnecker1993,Simon1995,Brandner1996,Ghez1997}.  One such survey in the Taurus-Auriga dark cloud found a companion frequency of 0.79$^{+0.12}_{-0.11}$ over all mass ratios and for separations of 3-5000 au for stars with primary masses of 0.25 - 0.7 M$_{\odot}$, roughly twice that of the Galactic field \citep{Kraus2011}.  In our recent survey of the M-star multiple population in the Orion Nebula Cluster (ONC), we found a companion frequency of 0.08$^{+0.04}_{-0.02}$ over mass ratios of 0.6 - 1.0 and separations of 30 - 160 au, consistent with the low-mass Galactic field population over the same parameter space \citep{DeFurio2019}, a result supported by \citet{Strampelli2020}.  Importantly, these star-forming regions have disparate present-day stellar number densities, 1-10 pc$^{-3}$ in Taurus \citep{Luhman2009ApJ...703..399L}, and 10$^{3.5-4.5}$ pc$^{-3}$ in the ONC \citep{HillenbrandHartmann,MarksKroupa2012} while also thought to have experienced different densities throughout their lifetimes \citep{Parker2014MNRAS.445.4037P}.

%These encounters, more likely in high density regions, can disrupt the multiple system either over the observed separations (30 - 160 au), or more likely at wider separations and earlier times, preventing an initially wide companion from migrating  to closer separations.  These interactions are significantly less likely to occur in low density associations where more multiple systems can survive and migrate to closer separations than observed in the ONC.

%Likely, low binding energy multiples like wide brown dwarf binaries can be broken up in the ONC over a few Myr, while higher binding energy binaries (smaller separation and higher mass ratio) survive and contain a trend of increasing power-law exponent with decreasing primary mass.

In \citet{DeFurio2019}, we estimated the companion frequency in the ONC over specific mass ratios and separations by identifying companions using a double point-spread function (PSF) fitting routine with empirical PSF models \citep{AndersonKing2006}. However, we were unable to place constraints on the mass ratio and orbital separation distributions due to the low number of detections.  This was in part because we analyzed archival Hubble Space Telescope (HST) data from the Advanced Camera for Surveys (ACS) from the HST Treasury Program of the ONC \citep[PID: 10246, ][]{Robberto2013} in only one filter (F555W).  This limitation reduced the sample size of our survey due to severe saturation in long exposures.  We updated our analysis routine to be sensitive to companions down to separations of 0.025" \citep{DeFurio2022ApJ...925..112D}, a region of parameter space in the ONC only explored for a handful of low-mass stars by \citet{Duchene2018}.  It is necessary to probe this parameter space given the distance to the ONC \citep[400 pc, ][]{grosschedl2018} because companions to M-type primary stars are commonly found around separations of 20 au in the Galactic field.

An expanded multiplicity survey with all available HST$/$ACS data is necessary to increase the sample size of the survey, identify more multiple systems, and derive specific functional forms of the companion population in terms of mass ratio and separation.  As low-mass M-type stars dominate the stellar population in terms of number and mass, placing constraints on its companion population in the ONC is crucial to understanding the dominant mechanisms of stellar multiple formation and identifying the impact of dynamics on multiple evolution.

In Section 2, we describe the data, the expanded ONC sample, and our method to detect companions.  In Section 3, we present our companion detections and characterization of the companion population. In Section 4, we compare our results to those of the Galactic field and Taurus.  In Section 5, we summarize our conclusions.

%In this paper, we present an expanded survey for companions to low-mass stars in all filters available in the HST Treasury Program ACS data (F435W, F555W, F658N, F775W, and F850LP).  We apply the companion detection technique developed in \citet{DeFurio2022ApJ...925..112D} to detect companions down to 0.025" (10 AU at 400 pc) in the best case scenario.  With this improved method, we will characterize the companion population of the ONC in terms of separation and mass ratio and look for 

% AND FLUX CUTOFF: with an unnecessarily low flux cutoff.  This limited 

%We detected 14 multiple systems overall, but some outside the common sensitivity of our method on the sample  (i.e. smaller separation or lower mass ratio).  This left only six binaries over which to estimate the companion frequency.  

\section{Methods} \label{sec:methods}

In order to identify companions to our sample of stars in the ONC, we first applied the double-PSF fitting technique described in \citet{DeFurio2019}, hereafter Paper I.  Then, we implemented the updated analysis, as described in \citet{DeFurio2022ApJ...925..112D} hereafter Paper II, in order to identify fainter and/or closer companions.  We used data from all five ACS filters used in the HST Treasury Program of the ONC (F435W, F555W, F658N, F775W, and F850LP) to identify companions to a target list described in Sec. \ref{subsec:subsample}.  This, combined with a Bayesian analysis of our results (described in Sec. \ref{sec:discussion}), increased the sample size by a factor of three compared to the single filter, frequentist approach of Paper I.

\subsection{The Data} \label{subsec:data}

We downloaded archival HST data taken with ACS on the Wide Field Camera (WFC) from GO program 10246 \citep[PI: M. Robberto][]{Robberto2013}. The data analyzed in this paper were obtained from the Mikulski Archive for Space Telescopes (MAST) at the Space Telescope Science Institute, and can be accessed via \dataset[DOI]{http://dx.doi.org/10.17909/5w65-rr60}.  This program captured a large mosaic of the ONC, covering 627 arcminutes, with four broadband filters and one narrow band filter: F435W, F555W, F658N, F775W, and F850LP.  Each of the 104 pointings had an integration time of 420s, 385s, 340s, 385s, and 385s respectively and a field of view of 202" x 200", with no dithering or CR-SPLITing.  The plate scale of ACS/WFC is 0.05"/pixel, undersampled across the visible spectrum (e.g. $\lambda$/D = 0.037" for F435W).  We used empirical PSFs to fit the data and accurately estimate the centroid of the primary PSF and any potential companion.  We use the $\_$flt data products from the HST pipeline, which have been bias-subtracted and flat-fielded.  These are the images specified by \citet{AndersonKing2006}, hereafter AK06, for which the empirical PSFs apply.  Within these long exposures, many images suffered from severe saturation which prevented us from performing PSF-fitting to many known members.  In Paper II, we showed that a 0.2 M$_{\odot}$ star is at the saturation limit in the F435W exposures (as is a $\sim$ 0.55 M$_{\odot}$ star with A$_{v}$ = 2), and a 0.03 M$_{\odot}$ object is at the saturation limit in the F850LP exposures (0.07 M$_{\odot}$ with A$_{v}$ = 3).  A more thorough description of the data can be found in \citet{Robberto2013}.

\subsection{Double-PSF Fitting with Empirical PSFs} \label{subsec:model}

We have described the double-PSF fitting process and the PSF models in Paper I, but provide a summary below.  The empirical PSFs were constructed by AK06 specifically for the ACS/WFC in multiple filters using real data across both detectors.  For each filter, they created a library of 90 PSFs, spread across both chips of the ACS, 4x super-sampled over a radius of 12.5 ACS/WFC pixels.  All filters used had a specific PSF library, except for F555W, and instead we use the F606W PSF library, as the differences are negligible (J. Anderson, private communication). To construct a PSF at a given location on the detector, we identify the four nearest empirical PSFs.  For each pixel within a 21x21 stamp centered on the target, we perform a bi-cubic interpolation of each of the four empirical PSFs based on the distance of that pixel to the center of the stamp and linearly interpolate those resulting four pixel values based on the proximity of each empirical PSF to the location of the target on the detector. This results in a detector-sampled PSF.  AK06 also provide a function to perturb the PSF models based on the brightest PSFs within a given image to account for changes in focus and pointing instability.  We applied the perturbation to the PSF library of each filter for each image, effectively generating slightly different PSFs for each observation.

In order to identify companions, we fit a double-PSF model (constructed from two detector-sampled PSFs) to a postage stamp of 21x21 pixels centered around a given target that has already had the mean background subtracted from an inner and outer annulus of 10 and 15 pixels, respectively.  Our model has six parameters: x and y center of the primary, combined magnitude of the system, separation between the centers of the primary and secondary, position angle corresponding to the center of the secondary, and the difference in magnitude ($\Delta$mag) between the primary and secondary.

We use the downhill simplex routine (AMOEBA) in IDL \citep{Press2007} to sample the parameter space and identify the best-fit six parameter model by minimizing the $\chi^{2}_{\nu}$ test statistic.  This process is repeated 200 times to adequately sample the parameter space and then derive errors for each variable.  Although we are deriving the best-fit double-PSF model to each input postage stamp, this does not mean that each fit is identifying a companion.  In Sec. \ref{subsec:dataanalysis}, we summarize how we differentiate between a true detection of a companion and a false positive fit.

\subsection{The Sample} \label{subsec:subsample}

The target list was constructed from \citet{DaRio2016}.  They used the Sloan 2.5m telescope and the APOGEE spectrograph (R $\sim$ 22,500) to obtain multi-object spectroscopy in the H-band of roughly 2700 sources in the ONC.  With these data, they were able to estimate the $\emph{T}_{eff}$, log $\emph{g}$, $\emph{v}$ sin$\emph{i}$, and extinction.

We searched through each of the 104 images in the five filters and identified which of the members exist within the data.  Then, we determined whether or not the central pixel value was above 55,000 counts, to exclude sources affected by saturation (AK06).  Finally, we ran the double-PSF fitting algorithm on each target in each image in which it was detected.  Following the prescription of Paper I and II, we set a chi squared threshold to indicate a reasonable fit where the data are likely described by the six-parameter double-PSF model, here a p-value of 0.1, $\chi^{2}_{\nu}$ = 1.774.  Nebulosity is the key factor for poor fits within these data as spatially variable extended emission is not factored into the model and results in high $\chi^{2}_{\nu}$ values. This process left us with 198 sources in F435W, 143 sources in F555W, 226 sources in F658N, 11 sources in F775W, and 5 sources in F850LP.  Source counts are very low in F775W and F850LP due to saturation.  In total, we have 276 unique sources throughout all filters, and 245 sources among the broadband filters with which we can estimate masses of the multiple systems.

\subsection{Companion Identification}\label{subsec:dataanalysis}

In Paper II, we thoroughly described the process by which we determine whether any given double-PSF fit is truly detecting a companion or is a false positive result.  We summarize this process below.

We generated 1000 artificial single stars for each of the various S/N values (15 - 130) and within each filter using the empirical PSF model libraries of AK06 across the detector with added Poisson noise.  Then, we ran the double-PSF fitting routine on each of these artificial single stars.  For every individual single star, we take each iteration from the fitting process (a six parameter binary model with a $\chi^{2}_{\nu}$ test statistic) and evaluate its chi squared probability P($\chi^{2}$).  We then defined a wide region of interest in the separation and $\Delta$mag parameter space where the code is likely to fit a binary model regardless of the true astrophysical scene.  This extends from 0 - 10 pixels in separation and 0 - 10 magnitudes in $\Delta$mag which encompasses the entirety of our search radius and continues beyond the background noise level.  This space was split into bins of 0.1 pixels and 0.1 magnitudes within which we select the best-fit, highest P($\chi^{2}_{\nu}$), model to each artificial single.  This process results in a probability distribution that describes the fit to each artificial single star.  We repeat this process for all artificial singles, and the resulting distributions are added giving a total probability distribution of fits to artificial singles of a given S/N and filter.

\begin{figure}[htb]
\gridline{\fig{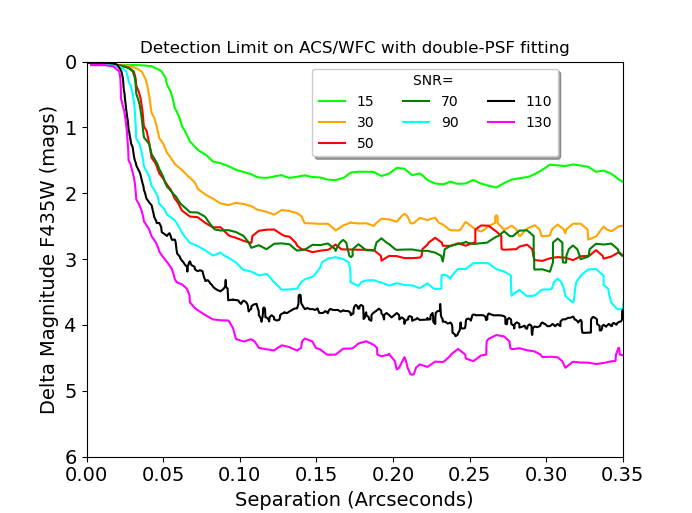}{0.5\textwidth}{}}
\caption{Each line shows the 0.1\% false positive probability line associated with our double-PSF fitting routine for the listed S/N in the F435W filter on HST/ACS, adapted from Paper II.}
\label{fig:435}
\end{figure}

These distributions represent the parameter space in which the code will converge around single stars, i.e. a false positive fit.  We sum the distribution vertically (in $\Delta$mag space) for each separation bin (0.1 pixels) which describes the $\Delta$mag where a certain percentage of the known false positive fits will lie below.  We can then take the normalized distribution of any binary fit to real data and multiply it by these false positive maps, based on the filter and S/N of the source, and sum the resulting distribution to arrive at the probability that a given binary fit to data is a false positive.  Based on our sample size, we define a false positive probability $<$ 0.1\% as indicative of a true binary detection, see Fig. \ref{fig:435}.

Our data set spans five filters within which specific sources can be imaged multiple times.  Within one filter, we can combine the false positive probability values derived for the multiple images of the same source if the fitted separation, $\Delta$mag, and position angle distributions are all consistent, i.e. the code is fitting the same feature within all images.  We evaluate this by calculating the Bhattacharyya coefficient (BC) between each pair of fits which determines the similarity between two distributions.  For distributions with BCs $>$ 0.1, we classify them as consistent and multiply the corresponding false positive probabilities to arrive at a total false positive probability within one filter.  For BCs $<$ 0.1, we use the best-fit among the images to calculate the false positive probability of the fit to this source.  For sources that appear in multiple filters, we calculate the BC of the separation and position angle distributions only as the $\Delta$mag is not expected to be the same across filters.  We then calculate a global false positive probability by multiplying the individual probabilities of the separate filters.  The resulting false positive probability is then evaluated as to whether it meets our criterion for a true binary detection ($<$ 0.1$\%$).

%\textcolor{red}{Two images or more: BC}

In Paper II, we also show that the completeness of our search entirely overlaps the region with false positive probability $<$ 0.1$\%$, meaning that we expect to be able to recover all companions that meet our detection criteria.  The completeness was determined by constructing artificial binaries in the same way we constructed artificial singles and evaluating how well we can recover the known parameters of these binaries through the $\chi^{2}_{\nu}$ test statistic.  We also compared the reliability of our false positive analysis from artificial single stars to the same process using real single stars.  We showed that the 0.1\% false positive probability curve is nearly identical between real and artificial singles, highlighting the accuracy of the empirical PSFs as models to real data.  Refer to Paper II for a more detailed description of this entire process with figures portraying these results.

\section{Results} \label{sec:results}

\subsection{Detections} \label{subsec:detections}
\begin{figure}[htb]
\gridline{\fig{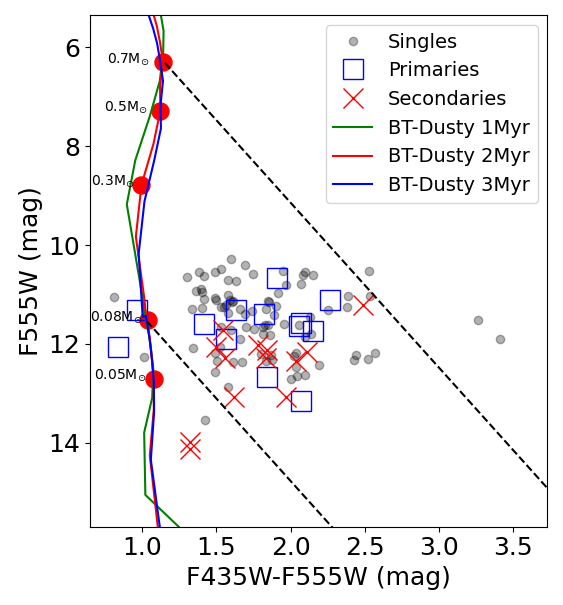}{0.47\textwidth}{}}
\caption{Color magnitude diagram for all the sources that were included in our sample with F435W and F555W data.  The detected binaries are plotted with both the primary and secondary components.  The solid lines are the BT-Dusty 1-3 Myr isochrones with specific masses chosen for reference.  The black dashed line represent the extinction vector derived from \citet{Cardelli1989}.}
\label{fig:colormag_diagram}
\end{figure}

\begin{figure*}[htb]
\gridline{\fig{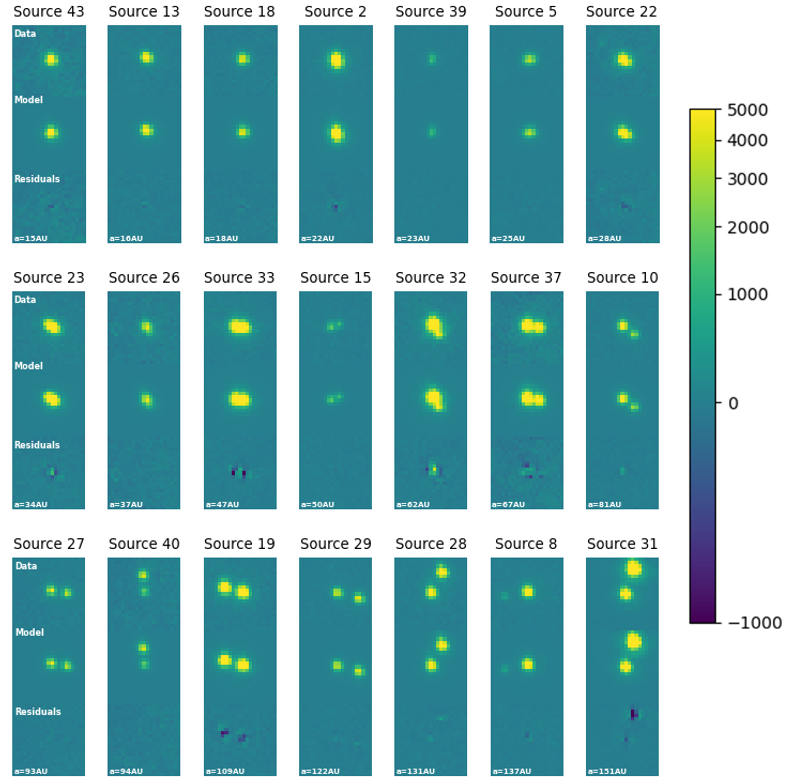}{\textwidth}{}}
\caption{We display all the binary detections made in the F435W filter, showing one image per detection. The top panel shows the 21x21 postage stamp cutout of the HST/ACS data, the middle panel shows our double-PSF model, and the bottom panel shows the residuals.  Each binary is labeled with their Source \# that corresponds to a 2MASS ID listed in Table \ref{table2} and the estimated separation is given at the bottom of each image.  Sources are listed left to right and top to bottom in order of increasing orbital separation.  The colorbar shows the dynamic range of all the images in units of detector counts.}
\label{fig:435binaries_1}
\end{figure*}

\begin{figure*}[htb]
\gridline{\fig{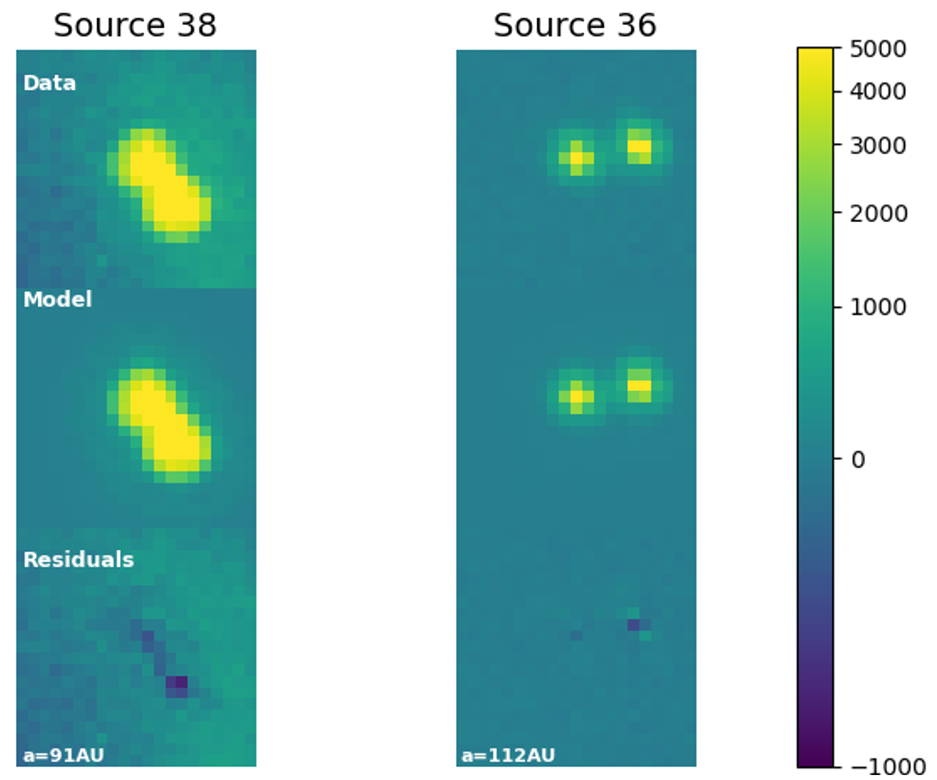}{0.7\textwidth}{}}
\caption{Same as Fig. \ref{fig:435binaries_1}  but for the F775W filter.}
\label{fig:775binaries_1}
\end{figure*}

We made 76 detections of 44 unique companions across all filters from the sample of 276 cluster members.  Twenty-one detections were made in the F435W filter, 24 in the F555W filter, 29 in the F658N filter, 2 in the F775W filter, and 0 in the F850LP filter.  Three detections were made only in the F435W filter, five in only F555W, ten in only F658N, and two in only F775W. Thirteen detections were made in both the F435W and F555W filters, 13 in both the F435W and F658N filters, 14 in both the F555W and F658N filters, and 8 in the F435W, F555W, and F658N filters. Thirty of the companions were previously detected \citep{Kohler2006,Reipurth2007,Robberto2013,Duchene2018,DeFurio2019,Strampelli2020}, five of which were only detected in our Paper I.  The redder filters had fewer detections due to saturation effects.  The F658N filter enabled the detection of the most companions because the narrow bandpass decreases the number of saturated sources.  In Table \ref{table1}, we present the photometry: the combined magnitude of the system and the difference in magnitude between the primary and companion.  We constructed a color-magnitude diagram (Fig. \ref{fig:colormag_diagram}) for all of the sources in our sample that were observed in the F435W and F555W filters, where both components of binary detections are shown. Figures \ref{fig:435binaries_1} - \ref{fig:775binaries_1} show an image of each detection made in every filter in which it was found.

The estimated projected separations range from 14 - 196 au (0.034"-0.49" or 0.68 - 9.8 ACS/WFC pixels) assuming a distance of 400 pc \citep{grosschedl2018}, and mass ratios range from 0.3 - 0.97.  We estimate masses using the isochrones produced from the BT-Dusty evolutionary models \citep{Chabrier2000,Allard2001,Allard2011, Allard2012} as they have updated opacities \citep{Barber2006MNRAS.368.1087B}.  We assume a cluster age of 2 Myr, the average age of the stellar population \citep{Reggiani2011}, and the extinction to each source as derived from \citet{DaRio2016}.  They used their estimated effective temperature of each source to derive the intrinsic \textit{(J-H)} color of the system and the observed \textit{(J-H)} 2MASS color to estimate the extinction, assuming a 2 Myr isochrone.  With these extinction values and the 2 Myr BT-Dusty isochrone, we estimate masses for the primary and companion of each binary system from the color information of Table \ref{table1}, assuming the same extinction to both components.  In Table \ref{table2}, we present the estimated physical parameters of each binary system along with essential values from the double-PSF fitting process.

\begin{figure}[htb]
\gridline{\fig{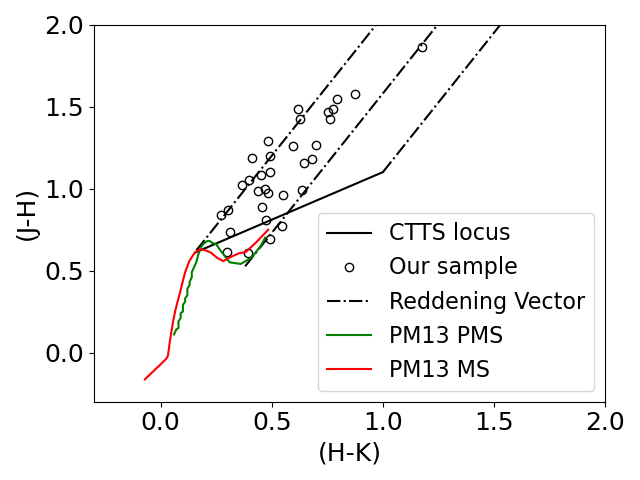}{0.45\textwidth}{}}
\caption{The J-H vs. H-K color-color plot for the sources with detected companions within our survey (black circles), showing no obvious signs of disk excess.  Over-plotted are the observed pre-main sequence (PMS) and main-sequence (MS) colors of \citet{Pecaut2013ApJS..208....9P}, PM13, the reddening vector \citep{Cardelli1989}, and the classical T-Tauri star (CTTS) locus from \citet{Meyer1997AJ....114..288M}.}
\label{fig:colorcolor}
\end{figure}

We investigated whether excess disk emission could impact the derivation of extinction for our binary detections.  We took the 2MASS JHK photometry and created a color-color plot comparing to the observed colors of known pre-main sequence stars \citep{Pecaut2013ApJS..208....9P}.  Significant disk emission would cause the observed near-IR photometry to be in excess of that expected from the photosphere, where the reddest photometry has a larger difference, causing motion toward the right in Fig. \ref{fig:colorcolor}.  With the inclusion of the reddening vector \citep{Cardelli1989} and the classical T-Tauri locus \citep{Meyer1997AJ....114..288M}, we show that there is no large near infrared excess observed.

Because sources were found in multiple filters, we used the data from the reddest filter in which they were found to estimate masses as excess emission from accretion shocks may impact blue/UV photometry \citep{Azevedo2006}.  We do not use data from the F658N (H$_{\alpha}$) filter to calculate masses as it will not be representative of the flux from the photosphere.

\begin{deluxetable*}{ccccccccc}
\tablenum{1}
\tablecaption{We list the observed magnitude information of each binary system.  In each set of two columns, we present the total magnitude of each system and the difference in magnitude between the primary and companion.  These values are the weighted mean calculated from all exposures of the target where the error is the 68\% confidence interval.  We use these values to estimate masses in Table \ref{table2}.  Typically, values are not listed due to saturation or low S/N in the particular filter, see Sec. \ref{subsec:subsample}.  For example, sources unsaturated in the long exposures of F435W or F555W are typically saturated in those of F775W as the peak of their flux occurs at bluer wavelengths. \label{table1}}

\tablewidth{0pt}
\tablehead{
\colhead{Source \#} & \colhead{Total Mag} & \colhead{$\Delta$ mag} & \colhead{Total Mag} & \colhead{$\Delta$ mag} & \colhead{Total Mag} & \colhead{$\Delta$ mag} & \colhead{Total Mag} & \colhead{$\Delta$ mag}\\
\colhead{} & \colhead{(F435W)} & \colhead{(F435W)}& \colhead{(F555W)} & \colhead{(F555W)}& \colhead{(F658N)} & \colhead{(F658N)} & \colhead{(F775W)} & \colhead{(F775W)}}
\decimalcolnumbers
\startdata
    1 & - & - & - & - & 16.96  $\pm$ 0.01 & 1.1 $\pm$ 0.1 & - & -\\
    2 & 19.49 $\pm$ 0.01 & 0.57 $\pm$ 0.15 & - & - & 15.97  $\pm$ 0.01 & 0.46 $\pm$ 0.15 & - & -\\
    3 & - & - & - & - & 15.39  $\pm$ 0.01 & 0.04 $\pm$ 0.04 & - & -\\
    4 & - & - & - & - & 15.84  $\pm$ 0.01 & 0.20 $\pm$ 0.02 & - & -\\
    5 & 21.32 $\pm$ 0.02 & 0.38 $\pm$ 0.15 & 19.18  $\pm$ 0.01 & 0.42 $\pm$ 0.04 & - & - & - & -\\
    6 & - & - & 20.79  $\pm$ 0.01 & 0.51 $\pm$ 0.3 & - & - & - & -\\
    7 & - & - & - & - & 16.32  $\pm$ 0.01 & 0.25 $\pm$ 0.02 & - & -\\
    8 & 20.51 $\pm$ 0.02 & 2.88 $\pm$ 0.04 & 18.63  $\pm$ 0.01 & 3.46 $\pm$ 0.01 & 15.99 $\pm$ 0.01 & 2.58 $\pm$ 0.02 & - & -\\
    9 & - & - & - & - & 14.91  $\pm$ 0.01 & 1.10 $\pm$ 0.01 & - & -\\
    10 & 20.60 $\pm$ 0.01 & 1.22 $\pm$ 0.03 & 19.42  $\pm$ 0.01 & 0.22 $\pm$ 0.02 & - & - & - & -\\
    11 & - & - & 18.61  $\pm$ 0.01 & 0.77 $\pm$ 0.15 & 16.46 $\pm$ 0.02 & 0.8 $\pm$ 0.15 & - & -\\
    12 & - & - & 21.56  $\pm$ 0.03 & 0.3 $\pm$ 0.2 & 18.20  $\pm$ 0.02 & 0.61 $\pm$ 0.09 & - & -\\
    13 & 20.56 $\pm$ 0.01 & 0.9 $\pm$ 0.3 & 18.96  $\pm$ 0.01 & 0.95 $\pm$ 0.3 & - & - & - & -\\
    14 & - & - & 20.45 $\pm$ 0.01 & 0.2 $\pm$ 0.2 & - & - & - & -\\
    15 & 22.00 $\pm$ 0.02 & 0.54 $\pm$ 0.06 & 20.11  $\pm$ 0.02 & 0.41 $\pm$ 0.02 & 17.67  $\pm$ 0.02 & 0.52 $\pm$ 0.03 & - & -\\
    16 & - & - & - & - & 14.65  $\pm$ 0.01 & 1.56 $\pm$ 0.01 & - & -\\
    17 & - & - & - & - & 15.44  $\pm$ 0.01 & 1.84 $\pm$ 0.02 & - & -\\
    18 & 21.06 $\pm$ 0.01 & 0.30 $\pm$ 0.14 & 19.08  $\pm$ 0.01 & 0.5 $\pm$ 0.2 & 16.46  $\pm$ 0.02 & 0.4 $\pm$ 0.2 & - & -\\
    19 & 19.15 $\pm$ 0.01 & 0.01 $\pm$ 0.01 & - & - & 15.33  $\pm$ 0.01 & 0.22 $\pm$ 0.01 & - & -\\
    20 & - & - & 21.74  $\pm$ 0.05 & 0.48 $\pm$ 0.08 & - & - & - & -\\
    21 & - & - & - & - & 15.62  $\pm$ 0.01 & 0.04 $\pm$ 0.02 & - & -\\
    22 & 19.92 $\pm$ 0.01 & 0.99 $\pm$ 0.09 & 18.76  $\pm$ 0.01 & 0.41 $\pm$ 0.03 & - & - & - & -\\
    23 & 19.81 $\pm$ 0.01 & 0.43 $\pm$ 0.04 & - & - & - & - & - & -\\
    24 & - & - & 19.48  $\pm$ 0.02 & 1.54 $\pm$ 0.15 & 16.53  $\pm$ 0.02 & 1.7 $\pm$ 0.3 & - & -\\
    25 & - & - & 20.36  $\pm$ 0.01 & 0.87 $\pm$ 0.03 & 17.91  $\pm$ 0.04 & 0.8 $\pm$ 0.15 & - & -\\
    26 & 20.98  $\pm$ 0.01 & 1.46 $\pm$ 0.11 & 19.20  $\pm$ 0.01 & 1.66 $\pm$ 0.05 & 16.71 $\pm$ 0.02 & 1.27 $\pm$ 0.07 & - & -\\
    27 & 20.89  $\pm$ 0.01 & 0.34 $\pm$ 0.02 & 19.22  $\pm$ 0.01 & 0.12 $\pm$ 0.02 & 16.83 $\pm$ 0.02 & 0.42 $\pm$ 0.02 & - & -\\
    28 & 19.63  $\pm$ 0.02 & 0.23 $\pm$ 0.02  & - & - & - & - & - & -\\
    29 & 20.79  $\pm$ 0.01 & 0.33 $\pm$ 0.03 & 18.43  $\pm$ 0.01 & 0.10 $\pm$ 0.01 & 16.40  $\pm$ 0.01 & 0.28 $\pm$ 0.02 & - & -\\
    30 & - & - & 20.45  $\pm$ 0.01 & 2.2 $\pm$ 0.1 & - & - & - & -\\
    31 & 18.64  $\pm$ 0.01 & 1.43 $\pm$ 0.01 & - & - & - & - & - & -\\
    32 & 18.87  $\pm$ 0.01 & 2.23 $\pm$ 0.02 & - & - & 15.45  $\pm$ 0.01 & 1.79 $\pm$ 0.02 & - & -\\
    33 & 19.17  $\pm$ 0.01 & 0.19 $\pm$ 0.02 & - & - & 15.49  $\pm$ 0.02 & 0.17 $\pm$ 0.02 & - & -\\
    34 & - & - & 22.12  $\pm$ 0.03 & 0.62 $\pm$ 0.06 & - & - & - & -\\
    35 & - & - & 21.36  $\pm$ 0.03 & 1.33 $\pm$ 0.03 & 18.41  $\pm$ 0.03 & 0.27 $\pm$ 0.04 & - & -\\
    36 & - & - & - & - & - & - & 19.49 $\pm$ 0.02 & 0.16 $\pm$ 0.02\\
    37 & 19.22  $\pm$ 0.01 & 0.86 $\pm$ 0.02 & - & - & 15.34  $\pm$ 0.02 & 0.94 $\pm$ 0.03 & - & -\\
    38 & - & - & - & - & - & - & 17.61 $\pm$ 0.01 & 0.27 $\pm$ 0.01\\
    39 & 22.53  $\pm$ 0.02 & 0.09 $\pm$ 0.07 & 20.76  $\pm$ 0.01 & 0.8 $\pm$ 0.2 & 18.11  $\pm$ 0.02 & 0.20 $\pm$ 0.08 & - & -\\
    40 & 21.24  $\pm$ 0.04 & 0.7 $\pm$ 0.2 & 19.19  $\pm$ 0.02 & 0.71 $\pm$ 0.03 & 16.92  $\pm$ 0.05 & 0.62 $\pm$ 0.08 & - & -\\
    41 & - & - & - & - & 17.08  $\pm$ 0.02 & 0.08 $\pm$ 0.03 & - & -\\
    42 & - & - & 20.51  $\pm$ 0.02 & 1.12 $\pm$ 0.03 & 17.83  $\pm$ 0.01 & 0.88 $\pm$ 0.04 & - & -\\
    43 & 20.51  $\pm$ 0.02 & 0.6 $\pm$ 0.2 & 19.06  $\pm$ 0.01 & 0.5 $\pm$ 0.3 & - & - & - & -\\
    44 & - & - & - & - & 15.45  $\pm$ 0.01 & 1.96 $\pm$ 0.02 & - & -\\
\enddata
\end{deluxetable*}

\begin{deluxetable*}{ccccccccccc}
\tablenum{2}
% and applied a correction factor of 1.16 to convert from projected separation to physical separation \citep{Dupuy2011}

\tablecaption{Candidate binaries with masses (M${_\odot}$), mass ratios (q), projected separations in arcseconds, physical separation in au, and position angles in degrees.  To estimate masses, we used the 2 Myr BT-Dusty isochrone and the A$_{v}$ estimates from \citet{DaRio2016}.  We assumed a distance of 400 pc \citep{grosschedl2018}.  Binary parameters  are the weighted mean calculated from all exposures of the target where the error is the 68\% confidence interval.  We also show the S/N of the target for the filter within which it is brightest along with the $\chi^{2}_{\nu}$ associated with that filter.  We do not show the S/N and $\chi^{2}_{\nu}$ within each filter, but they fulfill the requirements of Sec. \ref{subsec:subsample}. \label{table2}}

\tablewidth{0pt}
\tablehead{
\colhead{Source \#} & \colhead{2MASS ID} & \colhead{M$_{prim}$} & \colhead{M$_{sec}$} & \colhead{q} & \colhead{A$_{v}$} & \colhead{Projected Sep.} & \colhead{Physical Sep.}  &  \colhead{PA (deg)} & \colhead{$\chi^{2}_{\nu}$} &  \colhead{S/N}\\
\colhead{} & \colhead{} & \colhead{(M$_{\odot}$)} & \colhead{(M$_{\odot}$)} & \colhead{} & \colhead{(mag)} & \colhead{(arcseconds) } & \colhead{(AU) } & \colhead{(E of N)} &&}
\startdata
    1 & J05341202-0524196 & - & - & - & - & 0.05 $\pm$ 0.001 & 20.2 $\pm$ 0.4 & 254.91 & 1.19 & 164\\
    2 & J05342650-0523239 & 0.20 & 0.15 & 0.78 & 1.14 & 0.054 $\pm$ 0.001 & 21.68 $\pm$ 0.4 & 283.21& 1.37 & 370\\
    3 & J05342753-0528284 & - & - & - & - & 0.050 $\pm$ 0.002 & 20.0 $\pm$ 0.4 & 236.64 & 1.72 & 517\\
    4 & J05344083-0528095 & - & - & - & - & 0.210 $\pm$ 0.001 & 83.0 $\pm$ 0.4 & 45.22 & 1.63 & 350\\
    5 & J05344878-0517464 & 0.30 & 0.26 & 0.87 & 2.93 & 0.063 $\pm$ 0.002 & 25.2 $\pm$ 0.8 & 96.16 & 1.52 & 389\\
    6 & J05345009-0517121 & 0.14 & 0.10 & 0.72 & 2.68 & 0.034 $\pm$ 0.002 & 13.6 $\pm$ 0.8 & 294.44 & 0.92 & 100\\
    7 & J05345099-0517565 & - & - & - & - & 0.250 $\pm$ 0.001  & 100.0 $\pm$ 0.4 & 49.8 & 1.22 & 195\\
    8 & J05345120-0516549 & 0.78 & 0.24 & 0.30 & 4.7 & 0.346 $\pm$ 0.001 & 137.2 $\pm$ 0.4 & 91.09 & 1.52 & 554\\
    9 & J05345265-0529452 & - & - & - & - & 0.398 $\pm$ 0.001 & 159.0 $\pm$ 0.4 & 176.86  & 1.00 & 670\\
    10 & J05345483-0525125 & 0.14 & 0.12 & 0.88 & 1.48 & 0.202 $\pm$ 0.001 & 80.8 $\pm$ 0.4 & 323.6& 1.18 & 250\\
    11 & J05345683-0521363 & 0.81 & 0.63 & 0.78 & 5.16 & 0.048 $\pm$ 0.002 & 19.0 $\pm$ 0.6 & 101.49 & 1.27 & 414\\
    12 & J05350121-0521444 & 0.10 & 0.084 & 0.84 & 3.03 & 0.067 $\pm$ 0.003 & 27 $\pm$ 1 & 283.57  & 0.91 & 38\\
    13 & J05350160-0524101 & 0.09 & 0.06 & 0.67 & 0.0 & 0.042 $\pm$ 0.003 & 16 $\pm$ 1 & 196.84  & 1.42 & 307\\
    14 & J05350161-0533380 & 0.64 & 0.59 & 0.92 & 6.47 & 0.035 $\pm$ 0.003 & 14 $\pm$ 1 & 52.59  & 0.93 & 190\\
    15 & J05350207-0518226 & 0.28 & 0.24 & 0.86 & 3.66 & 0.125 $\pm$ 0.001 & 49.8 $\pm$ 0.3 & 107.5 & 1.22 & 148\\
    16 & J05350243-0520465 & - & - & - & - & 0.350 $\pm$ 0.001 & 140.2 $\pm$ 0.4 & 118.82 & 1.53 & 605\\
    17 & J05350274-0519444 & - & - & - & - & 0.292 $\pm$ 0.001 & 116.8 $\pm$ 0.4 & 241.55  & 1.07 & 301\\
    18 & J05350309-0522378 & 0.32 & 0.26 & 0.83 & 2.93 & 0.044 $\pm$ 0.002 & 17.5 $\pm$ 0.8 & 46.22 & 0.99 & 255\\
    19 & J05350476-0517421 & - & - & - & - & 0.270 $\pm$ 0.001 & 110.0 $\pm$ 0.2 & 196.11 & 0.97 & 373\\
    20 & J05350617-0522124 & 0.36 & 0.30 & 0.84 & 5.99 & 0.43 $\pm$ 0.005 & 172 $\pm$ 2 & 185.79 & 0.70 & 37\\
    21 & J05350642-0527048 & - & - & - & - & 0.177 $\pm$ 0.001 & 70.8 $\pm$ 0.4 & 148.61  & 1.37 & 264\\
    22 & J05350739-0525481 & 0.13 & 0.10 & 0.76 & 0.56 & 0.070 $\pm$ 0.001 & 28.0 $\pm$ 0.4 & 240.08 & 0.75 & 238\\
    23 & J05350985-0519339 & 0.79 & 0.69 & 0.87 & 5.29 & 0.085 $\pm$ 0.001 & 34.0 $\pm$ 0.4 & 325.1  & 1.35 & 163\\
    24 & J05351021-0523215 & 1.20 & 0.69 & 0.57 & 6.87 & 0.080 $\pm$ 0.005 & 32 $\pm$ 2 & 200.74  & 0.98 & 153\\
    25 & J05351188-0521032 & - & - & - & - & 0.440 $\pm$ 0.005  & 176 $\pm$ 2 & 110.15  & 0.98 & 84\\
    26 & J05351227-0520452 & 0.55 & 0.31 & 0.56 & 4.39 & 0.093 $\pm$ 0.002 & 37.2 $\pm$ 0.8 & 113.23  & 1.16 & 166\\
    27 & J05351270-0527106 & 0.20 & 0.19 & 0.95 & 2.1 & 0.233 $\pm$ 0.001 & 93.2 $\pm$ 0.2 & 230.86 & 0.99 & 265\\
    28 & J05351365-0528462 & 0.66 & 0.61 & 0.92 & 4.64 & 0.329 $\pm$ 0.001 & 131.5 $\pm$ 0.2 & 62.82  & 1.15 & 314\\
    29 & J05351445-0517254 & 0.50 & 0.49 & 0.98 & 3.85 & 0.297 $\pm$ 0.001 & 118.6 $\pm$ 0.4 & 255.47  & 1.45 & 664\\
    30 & J05351475-0534167 & 1.02 & 0.47 & 0.46 & 7.34 & 0.067 $\pm$ 0.003 & 27 $\pm$ 1 & 143.29  & 0.97 & 193\\
    31 & J05351491-0536391 & 0.40 & 0.24 & 0.58 & 1.88 & 0.387 $\pm$ 0.001 & 150.6 $\pm$ 0.4 & 73.38  & 1.50 & 690\\
    32 & J05351534-0519021 & 0.18 & 0.062 & 0.34 & 0.0 & 0.157 $\pm$ 0.001 & 63.0 $\pm$ 0.4 & 252.44 & 1.68 & 400\\
    33 & J05351547-0527227 & 0.59 & 0.55 & 0.94 & 3.87 & 0.117 $\pm$ 0.001 & 47.0 $\pm$ 0.4 & 218.2 & 1.69 & 366\\
    34 & J05351624-0528337 & 0.42 & 0.34 & 0.81 & 6.76 & 0.167 $\pm$ 0.003 & 67 $\pm$ 1 & 355.36  & 0.87 & 25\\
    35 & J05351676-0517167 & 0.13 & 0.064 & 0.51 & 2.86 & 0.265 $\pm$ 0.002 & 105.8 $\pm$ 0.8 & 118.05 & 0.97 & 75\\
    36 & J05351789-0518352 & - & - & - & - & 0.280 $\pm$ 0.001 & 111.8 $\pm$ 0.4 & 99.38  & 1.27 & 211\\
    37 & J05351794-0525338 & 0.36 & 0.26 & 0.72 & 2.26 & 0.167 $\pm$ 0.003 & 67 $\pm$ 1& 332.53 & 0.65 & 213\\
    38 & J05351884-0522229 & 0.04 & 0.03 & 0.9 & 0.0 & 0.226 $\pm$ 0.001 & 90.6 $\pm$ 0.4 & 124.72  & 1.25 & 449\\
    39 & J05351986-0531038 & 0.65 & 0.49 & 0.75 & 6.64 & 0.061 $\pm$ 0.001 & 24.2 $\pm$ 0.4 & 168.32  & 1.04 & 162\\
    40 & J05352017-0523085 & - & - & - & - & 0.232 $\pm$ 0.005  & 93 $\pm$ 2 & 194.32& 0.74 & 215\\
    41 & J05352032-0536394 & - & - & - & - & 0.099 $\pm$ 0.001 & 39.6 $\pm$ 0.2 & 324.78  & 1.68 & 179\\
    42 & J05352190-0515011 & 0.36 & 0.24 & 0.67 & 4.52 & 0.103 $\pm$ 0.001 & 41.4 $\pm$ 0.4 & 90.65  & 1.06 & 166\\
    43 & J05352534-0525295 & 0.16 & 0.12 & 0.77 & 1.24 & 0.035 $\pm$ 0.003 & 14 $\pm$ 1 & 89.78  & 1.02 & 203\\
    44 & J05352543-0521515 & - & - & - & - & 0.489 $\pm$ 0.001 & 195.6 $\pm$ 0.2 & 64.46 & 1.48 & 242\\
\enddata
\end{deluxetable*}

\subsection{Chance Alignments} \label{subsec:contaminants}
In order to perform a statistical analysis of our results, we must determine the probability that our candidates are not physically associated with their primaries, i.e. chance alignments.  This can occur from foreground stars, background stars, and other cluster members along the line of sight.  In Paper II, we used the photometric list of \citet{Robberto2013} to derive the stellar surface density as a function of radius from the cluster core that takes into account any object within the region regardless of its membership.  At a distance of 115" from the cluster core (as defined by distance from Theta 1 Ori C), the stellar surface density is 0.0034 stars/arcseconds$^{2}$ for the F435W and F555W filters, beyond which a vast majority of our sources for which we can estimate masses appear.  We expect to have 0.74 chance alignments give this stellar surface density, a search radius of 0.5", and 276 sources in total.

\subsection{Companion Population Analysis} \label{subsec:companionpopulation}
\begin{figure}[htb]
\gridline{\fig{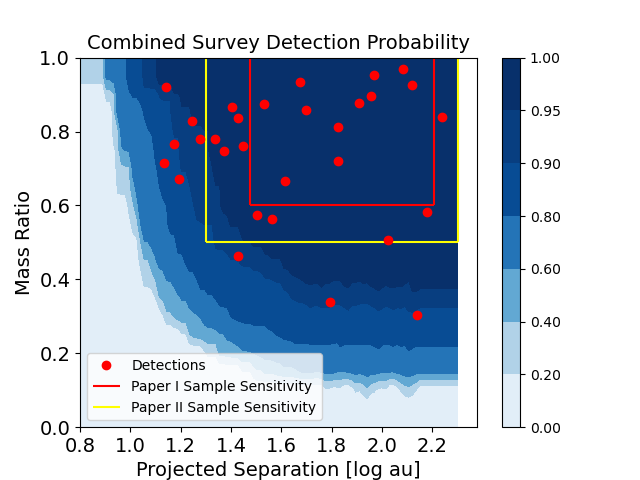}{0.5\textwidth}{}}
\caption{Summed detection probabilities for all the 245 sources in our survey in which we can calculate masses.  Red circles show the projected separation and estimated mass ratios for all detected companions.  Red and yellow lines show the detection limits from our past surveys in Paper I and II, respectively.  We use this map in Sec. \ref{subsec:companionpopulation} to model the separation and mass ratio distribution of the companion population in the ONC.}
\label{fig:survey_detetionprob}
\end{figure}

In Papers I \& II, we estimated the companion frequency of low-mass stars in the ONC by defining the region of mass ratio and projected orbital separation space over which we were complete for companions for most of our sample (a = 30 - 160 au and q = 0.6 - 1.0, and a = 20 - 200 au and q = 0.5 - 1.0, respectively), see Fig. \ref{fig:survey_detetionprob}.

To construct this sub-sample, we defined the detection limit for each source given the 0.1\% false positive probability curves associated with their S/N, e.g. Fig. \ref{fig:435}.  At each separation, there is an associated achievable contrast.  This contrast value is converted into a companion mass using the 2 Myr BT-Dusty isochrones, as described in Sec. \ref{subsec:detections}, and the minimum mass ratio that can be detected at each separation.  For sources that appear in multiple filters, we assign the mass ratio limit based on the lowest achievable mass ratio between filters.  We then sum over mass ratios and orbital separations to construct a combined detection probability of the entire sample, see Fig. \ref{fig:survey_detetionprob}, where each value of mass ratio and orbital separation has a corresponding fraction of the sample over which we are complete to companions.  In our frequentist approach of Paper I $\&$ II, we took this map and determined the mass ratio and separation range where we could detect companions to a vast majority of our sample in order to perform our statistical analyses.  In this paper, we use a Bayesian approach in order to preserve all information of the survey, and do not lose information in terms of sensitivity or true detections at close separations or low mass ratio.

We carry out a Bayesian analysis similar to the approach of \citet{Fontanive2018, Fontanive2019MNRAS.485.4967F} to fit models of the companion population to our data without the known biases of fitting binned histogram data \citep[e.g. ][]{allen2007ApJ...668..492A, Kraus2011, KrausHill2012}.  Bayes' theorem 

\begin{equation}
    P(\theta|D) \propto P(D|\theta)  P(\theta)
    \label{Bayes}
\end{equation}

is based on the concept that we can evaluate the likelihood function, P(D$\vert\theta$), given the observed data (D) and some initial model distribution or prior, P($\theta$), which then informs the model ($\theta$) and produces a posterior distribution P($\theta\vert$D) giving the probability that the model reproduces the data.  From the posterior distribution, we can then derive the values of the model parameters that best fit the data with their associated errors as this gives a probability density function (PDF) for each variable of the model.

We assume simple functional forms for the mass ratio and projected orbital separations distributions guided by previous work.  We define the mass ratio distribution as a power-law:

\begin{equation}
    \frac{dN_{1}}{dq} \propto 
    \begin{cases}
        q^{\beta} & \text{if } \beta \geq 0\\
        (1-q)^{-\beta} & \text{if } \beta < 0
    \end{cases}
    \label{qdistribution}
\end{equation}

where $\beta$ is the power-law index, and the piecewise function ensures symmetry about q=0.5 as defined in \citet{Fontanive2019MNRAS.485.4967F}.  We define the projected orbital separation distribution as a log-normal:

\begin{equation}
\frac{dN_{2}}{da} = \frac{1}{\sqrt{2\sigma_{loga}^{2}}} e^{-\frac{(log(a)-log(a_{o}))^{2}}{2\sigma_{loga}^{2}}}
    \label{adistribution}
\end{equation}

where log(a$_{o}$) is the mean and $\sigma_{loga}$ is the standard deviation of the distribution.

We use the python module PyMultiNest \citep{Buchner2014A&A...564A.125B} that performs the Nested Sampling Monte Carlo analysis using MultiNest \citep{Feroz2009MNRAS.398.1601F} to adequately sample our parameter space and estimate the best fitting model to our observations.  Importantly, PyMultiNest also calculates the evidence of the model tested by integrating eq. \ref{Bayes} over all parameters which can then be used to compare different models.  Nested sampling is based on generating a user-defined large number of live points that are sampled directly from the prior distribution with their likelihoods evaluated and sorted.  Then, points are sampled again from the priors until a point reaches a likelihood greater than the lowest likelihood of the original set of live points.  This process continues to maximize the likelihood until the effect of sampling becomes negligible.  \citet{Buchner2014A&A...564A.125B} describe this as a "scan vertically from the least probable zones to the most probable".

We define our likelihood function used in the sampling routine based on Poisson statistics and the physical parameters of our detected companions \citep{Fontanive2018}.  First, the Poisson likelihood is calculated with:

\begin{equation}
\mathcal{L}_{p} = \frac{k^{d} e^{-k}}{d!}
    \label{poissonlike}
\end{equation}

where \textit{k} (mean of the Poisson distribution) is the expected number of companion detections given the model and \textit{d} is the number of detections observed in the data.  We evaluate the expected number of detections with 

\begin{equation}
k = \sum_{i=1}^{n} p_{i}*CF*\frac{N}{n}
    \label{expected}
\end{equation}

where \textit{CF} is the companion frequency of the sample (mean number of companions per primary star) over all mass ratios and 10-200 au (our survey sensitivity in projected separation), \textit{N} is the total number of sources in our sample (here, 245), \textit{n} is the number of generated companions in the sampling process, and \textit{p$_{i}$} is the probability that the \textit{i} generated companion will be detected given our survey sensitivity.  We assign \textit{p$_{i}$} based on the physical parameters of the generated companions and the probability that they would be detected in our survey, see Fig. \ref{fig:survey_detetionprob}.  A companion generated with q = 0.01 and log(a) = 1.0 would result in \textit{p$_{i}$} = 0.0, but a companion with q = 1.0 and log(a) = 1.8 would result in \textit{p$_{i}$} = 1.0.

Each time we evaluate the likelihood of a model, we sample the prior distribution for $\beta$, log(a$_{o}$), $\sigma_{loga}$, and \textit{CF}.  For our case, we do not have informed priors on the free parameters, so we generate flat distributions with -5.0 $<$ $\beta$ $<$ 5.0, 0.0 $<$ log(a$_{o}$) $<$ 4.0, 0.1 $<$ $\sigma_{loga}$ $<$ 5.0, and 0.0 $<$ \textit{CF} $<$ 1.0.  Priors for log(a$_{o}$) and $\sigma_{loga}$ are log-flat, so that each au is equally weighted.  Then, we generate the mass ratio and projected orbital separation distributions of eqs. \ref{qdistribution} and \ref{adistribution} based on those sampled values.  From each model, we generate n = 10$^{4}$ companions, with mass ratios = 0 - 1 and projected orbital separations = 10 - 200 au, and determine the detection probability \textit{p$_{i}$} for each based on Fig. \ref{fig:survey_detetionprob}.  We sum over all \textit{n} generated companions to derive the expected number of detections \textit{k}, and then compute the Poisson likelihood where \textit{CF} is a free parameter.

We then calculate the total likelihood given the information from the model and our real detections.  As done in \citet{Fontanive2018}, we use the sampled parameters to generate the model.  We then multiply the model by the combined survey detection probability in mass ratio and projected orbital separation space.  This produces a joint distribution that gives the expected companion distribution based on the sensitivity of the survey.  Then, we calculate the joint probability of each true detection from our survey given that sampled model distribution and our detection limits, \textit{p$_{j}$}.  The total likelihood is then calculated as:

\begin{equation}
\mathcal{L} = \mathcal{L}_{p} * \prod_{j=1}^{d} p_{j}
    \label{likelihood}
\end{equation}

We use our list of detected multiple systems with estimated primary masses of 0.08-0.7 M$_{\odot}$ (24 in total) to represent the low-mass stellar population as the companion properties can depend on primary mass.  Stars with masses $>$ 0.7 M$_{\odot}$ were only observable in F435W or F555W due to their high extinctions, typically A$_{v}$ $>$ 5 mag.

\begin{figure*}[htb]
\gridline{\fig{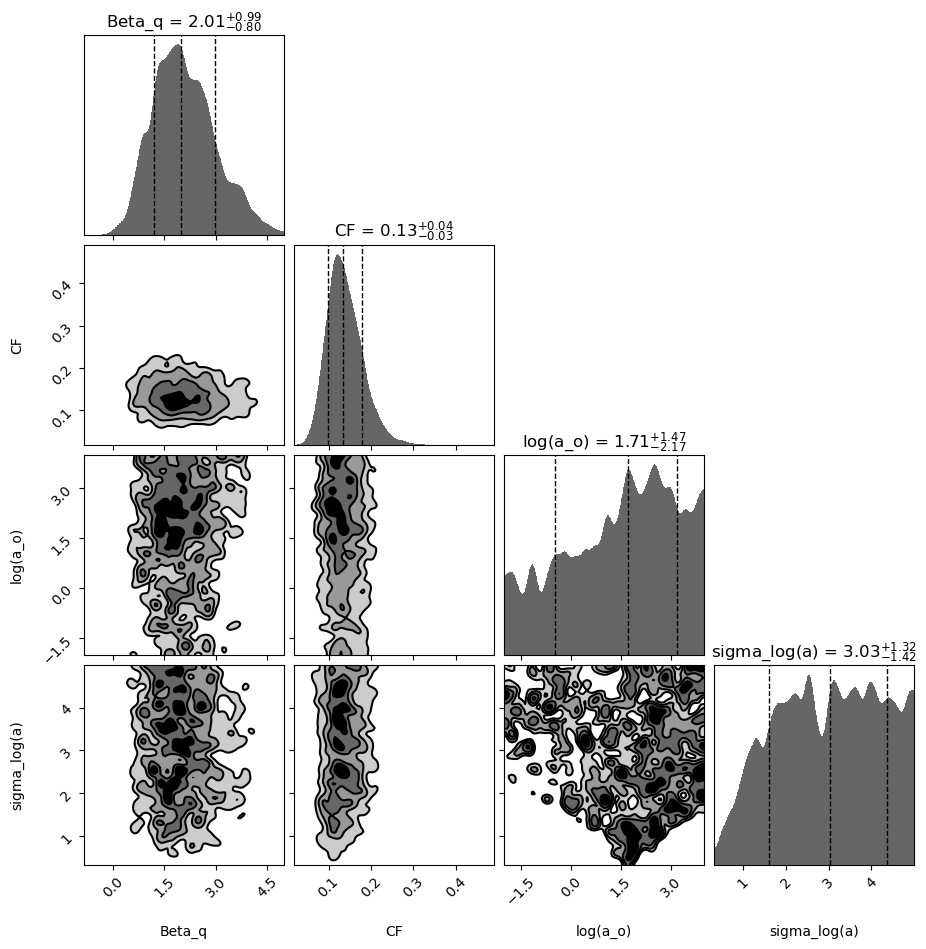}{0.5\textwidth}{}
    \fig{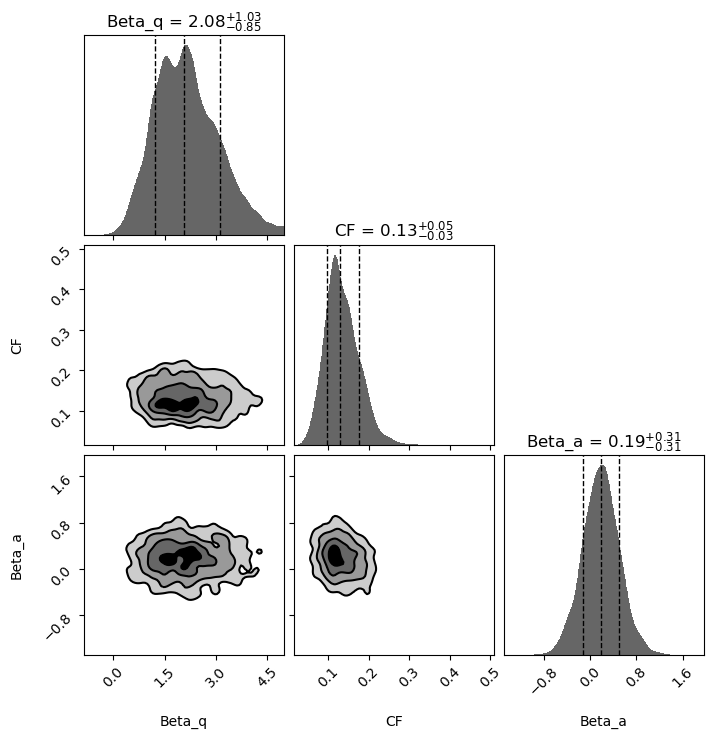}{0.5\textwidth}{}}
\caption{Corner plots representing the posterior distributions of the four and three parameter models used in our fit to the companion population.  Beta\_q is the exponent to the power law model of the mass ratio distribution, CF is the companion frequency over q=0-1 and a=10-200 au, log(a\_o) is the mean of the log-normal separation distribution used in the four parameter model, sigma\_log(a) is the standard deviation of the log-normal separation distribution used in the four parameter model, and Beta\_a is the power law exponent to the separation distribution used in the three parameter model.  The separation distribution is unconstrained with a log-normal, likely due to the limited range of our survey.  Evidence values show no statistical preference between the 3 and 4 parameter model, and therefore we use the three parameter model fits for further analysis.}
\label{fig:cornerplots}
\end{figure*}

From our sample of 24 low-mass binaries in the ONC, we estimated the following four parameters over all mass ratios and projected orbital separation 10-200 au with 1$\sigma$ errors (68\% confidence interval): $\beta$ = 2.01$^{+0.99}_{-0.80}$, \textit{CF} = 0.13$^{+0.04}_{-0.03}$, log(a$_{o}$) = 1.71$^{+1.47}_{-2.17}$, and $\sigma_{loga}$ = 3.03$^{+1.32}_{-1.42}$. As shown in Fig. \ref{fig:cornerplots}, the parameters of the projected orbital separation distribution are unconstrained, likely due to our sensitivity between 10 - 200 au (only 1.3 orders of magnitude).  Unless the distribution was very narrow, this search radius limited our ability to fit a log-normal distribution to the detected companions.

We then re-ran our fitting code to model the separation distribution as a power-law with index $\beta_{a}$, following eq. \ref{qdistribution}.  We estimated the following three parameters with 1$\sigma$ errors: $\beta$ = 2.08$^{+1.03}_{-0.85}$, \textit{CF} = 0.13$^{+0.05}_{-0.03}$, $\beta_{a}$ = 0.19$^{+0.31}_{-0.31}$, see Fig. \ref{fig:cornerplots}.  The estimated mass ratio power law index and companion frequency were consistent between both models.

One of the useful features of nested sampling is calculating the Bayesian evidence of a particular model to which other models can be compared.  The log-evidence of the four parameter model is -177.1 while the log-evidence of the three parameter model is -177.8.  \citet{Trotta2008ConPh..49...71T} show that when comparing two models a difference in the log-Bayesian evidence $<$ 1 is inconclusive and therefore no distinction can be drawn between the two models.

\section{Discussion} \label{sec:discussion}

Many studies in the Galactic field and in star-forming regions have characterized their companion population and estimated the companion frequency as a function of primary mass.  We can compare our results to other surveys in order to search for differences as a function of primary mass and star-forming environment which could inform our understanding of multiple formation and evolution.

\subsection{Comparison to the Galactic Field} \label{subsec:fieldcomparison}
\subsubsection{Mass Ratio Distribution} \label{subsubsec:qcomparison}

\citet{ReggianiMeyer2013} analyzed two Galactic field multiplicity surveys to characterize the mass ratio distributions of solar-type \citep{Raghavan2010} and M-type \citep{Janson2012} stars.  They found the distributions of mass ratios between the two surveys consistent, and therefore combined the data and fit a power-law using the maximum-likelihood estimation of \citet{Feigelson2012msma.book.....F}, resulting in $\beta$ = 0.25 $\pm$ 0.29.  Using the results of our 3-parameter model, we see a 2.0$\sigma$ difference between our fitted power-law exponent and that of \citet{ReggianiMeyer2013}.  

The sample of primary stars in the \citet{Janson2012} survey to which they detected companions had a significant portion of masses $>$ 0.3 M$_{\odot}$.  A significant portion of the sources in our sample however have a mass $<$ 0.3 M$_{\odot}$, see Fig. \ref{fig:colormag_diagram}, potentially indicative of a difference in the mass ratio distribution of mid-late M-stars and early M-stars.  \citet{Offner2022arXiv220310066O} performed a similar analysis as \citet{ReggianiMeyer2013} where they fit a power-law to the mass ratio distribution of the results from the \citet{Winters2019AJ....157..216W} field M-star survey.  For primary masses 0.3 - 0.6 M$_{\odot}$, they find a power-law index of 0.1 $\pm$ 0.4, also a 2.1$\sigma$ difference compared to our model.  However, they also found a power-law index of 0.7 $\pm$ 0.5 for primary masses 0.15 - 0.3 M$_{\odot}$ (more similar to our sample, a difference of 1.4$\sigma$)

We can also compare our results to the brown dwarf multiplicity study of \citet{Fontanive2018} who found a power law index of 6.1$^{+4.0}_{-2.7}$ to their companion population.  This results in a 1.7$\sigma$ difference compared to our survey.

\subsubsection{Companion Frequency} \label{subsubsec:cfcomparison}

In our survey, we were also able to constrain the companion frequency of our sample over mass ratios = 0 - 1 and projected separations 10 - 200 au.  We can compare these results to those of \citet{Susemiehl2022A&A...657A..48S} who constrain the orbital separation distribution and companion frequency of M-stars in the Galactic field using a combination of multiple surveys that extend to primary masses of 0.13 M$_{\odot}$, more similar to our sample see Fig. \ref{fig:colormag_diagram}.  We take eqs. \ref{qdistribution} and \ref{adistribution} to define a new equation to characterize the companion frequency of a survey given the detection limits:

\begin{equation}
    CF = C_{n}*\int_{q_1}^{q_2} \frac{dN_{1}}{dq} \int_{a_1}^{a_2} \frac{dN_{2}}{da}
    \label{cf}
\end{equation}

where C$_{n}$ is an integration coefficient and CF is the companion frequency.

\citet{Susemiehl2022A&A...657A..48S} model the semi-major axis distribution as a log-normal resulting in a peak log(a$_{o}$) = 1.68$^{+0.14}_{-0.16}$ and $\sigma_{loga}$ = 0.97$\pm$ 0.19, having corrected projected separations to physical separations using a multiplicative factor of 1.26 \citep{Fischer1992ApJ...396..178F}.  They estimate the companion frequency over 0.6 $\leq$ q $\leq$ 1.0 and 0 $\leq$ a $\leq$ 10,000 au assuming the mass ratio distribution from \citet{ReggianiMeyer2013}, resulting in CF= 0.229 $\pm$ 0.028. Taking those bounds as the bounds of the integrals in eq. \ref{cf}, we can solve for the constant C$_{n}$.  We then integrate over the bounds of our own results (0.0 $\leq$ q $\leq$ 1.0 and 12.6 $\leq$ a $\leq$ 252 au after converting to physical separations with their same multiplicative factor) and arrive at an expected companion frequency of the Galactic field over the sensitivity of our survey.  We repeated this process 10$^{4}$ times, sampling the companion parameters given their errors from \citet{ReggianiMeyer2013} and \citet{Susemiehl2022A&A...657A..48S}, evaluating the expected companion frequency each time.  We estimate a companion frequency of 0.244$^{+0.076}_{-0.053}$ over mass ratios 0-1.0 and semi-major axes 12.6-252 au, 1.5$\sigma$ difference compared to our estimate in the ONC.

% use MLE value or median??
\subsection{Comparison to Taurus} \label{subsec:clustercomparison}
\citet{Kraus2011} performed a multiplicity survey in Taurus, sensitive to companions between 3-5000 au.  In order to directly compare our results to their companion population, we must restrict their sample of primary stars over a similar mass range and restrict the companion detections over a common separation and mass ratio to which our survey and theirs is sensitive.   Therefore, we restricted their sample to sources with primary masses $\leq$ 0.45 M$_{\odot}$, where the mean primary mass of the sample is 0.31 M$_{\odot}$ compared to 0.29 M$_{\odot}$ in our ONC sample.  Our Bayesian analysis evaluates the companion frequency over all mass ratios and projected separations = 10-200 au.  Of the 34 stars in their sample with masses $<$ 0.45 M$_{\odot}$, 12 had companions between projected separations of 10-200 au.  To determine how this compares to our survey, we converted these results to a binomial distribution following the formalism of \citet{Burg2003}.  Then, we sampled the results of our companion frequency distribution and integrated the binomial distribution of the results of \citet{Kraus2011} from 0 to the sampled ONC companion frequency to determine the posterior probability that our ONC results resemble the observed Taurus results.  Taking the mean of all the samples, we arrive at a probability of 0.002 that our ONC model can describe the Taurus observations.

\subsection{Implications} \label{subsec:implications}
Our analysis of the companion population of low-mass M-stars in the ONC identified 44 companions and constrained the mass ratio distribution with a power law fit with exponent $\beta$ = 2.08$^{+1.03}_{-0.85}$ and a companion frequency  = 0.13$^{+0.05}_{-0.03}$ over 10-200 au and q=0-1.  Due to our limited sensitivity in separation, we could not place meaningful constraints on the separation distribution.  

We identified a 2$\sigma$ difference between the power-law exponent of the mass ratio distribution for stars with masses $\geq$ 0.3 M$_{\odot}$ in the Galactic field from the studies of \citet{ReggianiMeyer2013} and \citet{Offner2022arXiv220310066O} and the results of our Bayesian analysis in the ONC for primary stars down to the hydrogen burning limit.  Importantly, we find a 1.4$\sigma$ difference between the mass ratio distribution of our results and those of \citet{Offner2022arXiv220310066O} for primary mass 0.15-0.3 M$_{\odot}$ and a 1.7$\sigma$ difference compared to the results of \citet{Fontanive2018} for brown dwarf primaries (likely due to their large errors).

\citet{KrausHill2012} searched for companions at separations = 7 - 5000 au to stars with masses 0.07 - 0.5 M$_{\odot}$ in Taurus and Upper Sco, and found a trend in the mass ratio distribution similar to the field where $\beta$ = 0.18$^{+0.33}_{-0.3}$ for 0.3 - 0.5 M$_{\odot}$ primaries and $\beta$ = 1.02$^{+0.59}_{-0.52}$ for 0.15 - 0.3 M$_{\odot}$ primaries.  Although these results are over a wider range of separations, the mass ratio distribution appears consistent with the Galactic field population for the same primary masses and also consistent with our results in the ONC for lower mass primaries (1$\sigma$ difference).  These results are potentially indicative of a primordial mass ratio distribution that is an outcome of the star formation process and mostly unaffected by dynamical interactions \citep{Parker2013MNRAS.432.2378P}, over the separations sampled.

Previous studies have identified preferential circumbinary disk accretion onto the companion as a mechanism to drive mass ratios toward unity \citep{YoungClarke2015MNRAS.452.3085Y, Satsuka2017MNRAS.465..986S}. These companions can be produced through turbulent fragmentation or disk fragmentation, and can migrate inward through interactions with the circumstellar disk or gas from the natal cloud, producing separation like those seen in our survey \citep{Bate2002MNRAS.336..705B, Bate2003MNRAS.339..577B, Offner2010}. \citet{Bate2012MNRAS.419.3115B} simulated a turbulent molecular cloud using a radiation hydrodynamical simulation, and found the distribution of mass ratios to stars with masses 0.1-0.5 M$_{\odot}$ weighted towards equal masses with 63\% of these multiple systems with q $>$ 0.6, and only 50\% for stars $>$ 0.5 M$_{\odot}$.  These processes may be applicable to our observations within the ONC as we see a mass ratio distribution weighted towards equal masses for low-mass primaries.

%Potentially add??
%\citet{KrausHill2012} searched for companions at separations = 7 - 5000 AU to stars with masses 0.07 - 0.5 M$_{\odot}$ in Taurus and Upper Sco, and found a similar trend in the mass ratio distribution where $\gamma$ = 0.18 for 0.3 - 0.5 M$_{\odot}$ primaries, $\gamma$ = 1.0 for 0.15 - 0.3 M$_{\odot}$ primaries, and $\gamma$ = 0.96 for 0.07 - 0.15 M$_{\odot}$ primaries. For the lowest mass primaries in that study, there is a distinct dependence on separation where $\gamma$ = 4.0 for companions $<$ 20 AU (separations of the vast majority of companions to field brown dwarfs) and $\gamma$ = -0.3 for companions $>$ 20 AU. Similarly, \citet{DeFurio2022ApJ...925..112D} found that the mass ratio distribution of 0.012 - 0.1 M$_{\odot}$ primaries with separations $>$ 20 AU in the ONC has a low probability of coming from the field brown dwarf mass ratio distribution \citep[$\gamma$ = 3.6,][]{Reid2006}, but found no significant difference compared to a flat mass ratio distribution.  These studies identify potential dependencies of the companion population on primary mass, separation, and birth environment.

In addition, we do not find a difference between the companion frequency of the ONC and that of the Galactic field \citep{Susemiehl2022A&A...657A..48S} for low-mass stars over q=0-1.0 and semi-major axes = 12.6-252 au.  However, we find a probability of 0.002 that the companion frequency of the ONC can describe the observations of the low-mass T-association Taurus-Auriga.

High stellar density has been linked to lower companion frequency due to increased dynamical interactions that can disrupt a multiple system \citep{kroupa1995, Kroupa2001}.  The ONC is a high density star-forming region, n $\sim$ 10$^{3.5-4.5}$ stars pc$^{-3}$ \citep{HillenbrandHartmann,MarksKroupa2012}, that is thought to have experienced a cool-collapse phase which temporarily increases the density by roughly a factor of 10 followed by relaxation and expansion \citep{Allison2009,Allison2010,ParkerGoodwinAllison2011}.  Within temporary states of higher density, dynamical interactions become significant over 10-200 au and could have disrupted many stellar multiple systems resulting in the present day configuration of the companion population differing from the initial state.  Based on the present day density of the ONC, most stellar multiple systems with separations of 10-200 au are not expected to disrupted in many 10s of Myr \citep{Weinberg1987, DeFurio2022ApJ...925..112D}.

Taurus-Auriga, is a low-mass, T-association, that has a low stellar density of 1-10 stars pc$^{-3}$ \citep{MarksKroupa2012, King2012MNRAS.427.2636K, Parker2014MNRAS.445.4037P}.  This region is not expected to have had a period of intense dynamical interactions, leaving the present day companion population likely very similar to the initial population.  Additionally, the current state of the low-mass star companion population with separations $\lesssim$ 200 au will remain unchanged in the future, in excess of the Galactic field population.

Higher density regions like Westerlund 1 (10$^{5}$ M$_{\odot}$ pc$^{-3}$) and Arches \citep[10$^{5.5}$ M$_{\odot}$ pc$^{-3}$, ][]{Clark}, would experience an even higher level of dynamical processing, likely destroying many multiple systems with separations between 10-200 au.  All types of star-forming regions will contribute to the Galactic field population, potentially with an equal number of stars over order of magnitude bins in total cluster stellar mass \citep{Gieles2009MNRAS.394.2113G}.  If the star-forming regions within the same stellar mass bin experienced roughly equal density states, it is possible that regions that do not experience very high density states (like Taurus) will contribute an excess of companions, ONC-like clusters will contribute a comparable number of companions, and even higher density regions like Arches will contribute a paucity of companions relative to the Galactic field population over these orbital separations.  In this case, the ONC would be representative of a companion population averaged over all star-forming regions commensurate with the Galactic field population.

For a more thorough analysis on the role of environment and primary mass on the formation and evolution of companion populations, we require wide-field imaging programs in diverse star-forming environments that are sensitive to members across a wide range of primary masses. The \textit{James Webb Space Telescope (JWST)} is capable of wide-field, diffraction-limited imaging comparable to HST in the near and thermal infrared.  GTO programs 1190, 1202, 1229, and 1256 will all have NIRCam or NIRISS imaging data in star-forming regions of various environments, NGC 2024, NGC 1333, IC 348, and the ONC respectively.  Our HST Cycle 30 program (GO-17141) is designed to perform a similar analysis as this paper (for primary masses 0.01-1.0 M$_{\odot}$) in NGC 1333, a moderate density star-forming region, and is well complemented by the overlapping JWST NIRISS program 1202.  The data from these programs will allow us to probe similar separations as this paper, and directly compare the demographics of the companion populations within regions of different densities and evolutionary states.  The up-the-ramp readout of infrared detectors on \textit{JWST} will also drastically reduce saturation effects like those seen on ACS/WFC.  These programs will also provide broad wavelength coverage in multiple filters, where the sensitivity in the infrared is also conducive to detecting members at the low-mass end of the initial mass function as well as very low mass ratios.

\section{Conclusion} \label{sec:conclusion}
We performed a multiplicity survey of low-mass stars in the ONC using a double-PSF fitting routine with empirical PSFs with archival \textit{HST/ACS} data.  This work expanded on our previous study by using all data available in five ACS filters, updating the analysis routine to be sensitive to fainter/closer companions, and applying a Bayesian analysis to our observations to characterize the companion population.  To summarize the results of our survey:

\vspace{0.35cm}
1) We detected 44 companions down to separations of 0.034" out of the 276 members in our sample, many of which were found in multiple filters. Thirty of the 44 were previously detected in other multiplicity surveys, five of which came exclusively from \citet{DeFurio2019}.

\vspace{0.35cm}
2) Using a Bayesian approach, we estimate the companion frequency as 0.13$^{+0.05}_{-0.03}$ for low-mass stars in the ONC and the exponent of the power law to the mass ratio distribution as 2.08$^{+1.03}_{-0.85}$ over projected separations of 10-200 au and all mass ratios.

\vspace{0.35cm}
3) We find a 1.4$\sigma$ difference in the mass ratio distribution power law exponent derived for low-mass stars in the ONC (2.08$^{+1.03}_{-0.85}$) compared to low-mass primaries (0.15 - 0.3 M$_{\odot}$) in the field (0.7 $\pm$ 0.5), a result consistent with that identified for low-mass primaries in Taurus and Upper Sco \citep[1.02$^{+0.59}_{-0.52}$,][]{KrausHill2012}.  These results are supportive of the hypothesis that the mass ratio distribution is mostly unaffected by dynamical processes and is the result of the star formation process itself \citep{Parker2013MNRAS.432.2378P}, cf. companion frequency.

\vspace{0.35cm}
4) We find that the companion frequency of low-mass stars in the Galactic field is consistent with that of the ONC.  However, we find a significant excess in Taurus relative to the ONC, with a probability of 0.002 that the companion frequency of the ONC can describe the observations in Taurus.

\vspace{0.35cm}
5) Our findings suggest that early dynamical processing of multiple systems in high density star-forming regions is important in sculpting the companion frequency, and that the ONC may be a more representative star-forming region that will contribute to the Galactic field.

\begin{acknowledgments}
We would like to thank Jay Anderson for many productive discussions on PSF modeling and the implementation of his PSF code, as well as Megan Kiminki for contributions to the construction of our code. This work is based on observations made with the NASA/ESA Hubble Space Telescope, obtained from the data archive at the Space Telescope Science Institute. STScI is operated by the Association of Universities for Research in Astronomy, Inc. under NASA contract NAS 5-26555. Support for this work was provided by NASA through grant number HST-AR-15047.001-A from the Space Telescope Science Institute, which is operated by AURA, Inc., under NASA contract NAS 5-26555. For the purpose of open access, the author has applied a Creative Commons Attribution (CC BY) licence to any Author Accepted Manuscript version arising from this submission.
\end{acknowledgments}

\bibliography{draft}{}
\bibliographystyle{aasjournal}

\begin{figure*}[htb]
\gridline{\fig{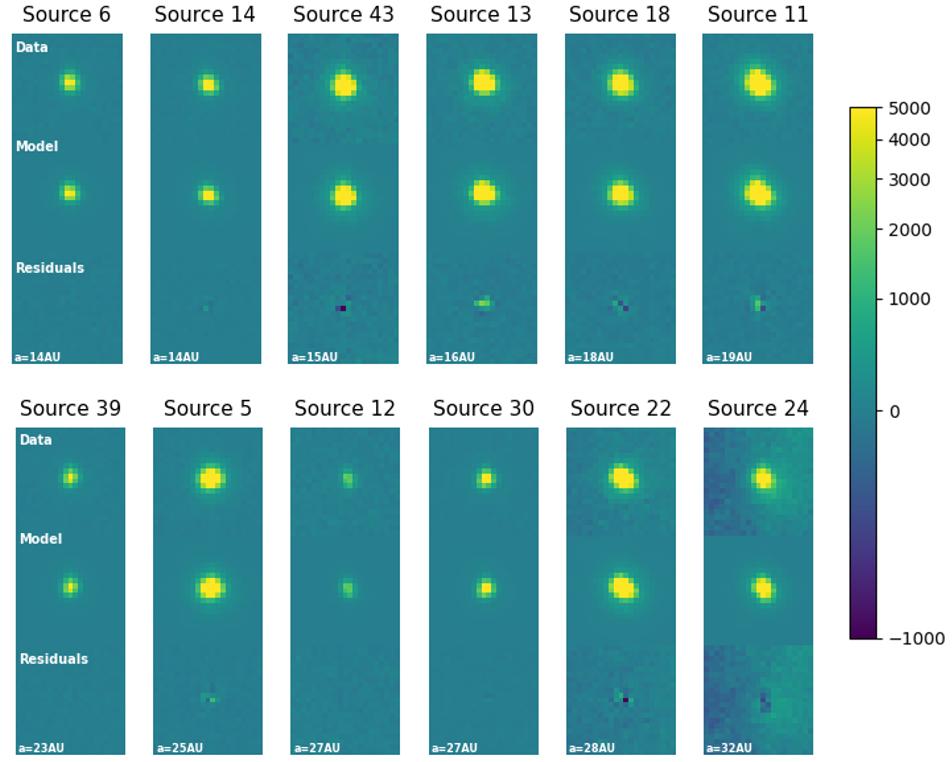}{\textwidth}{}}
\caption{Same as Fig. \ref{fig:435binaries_1}  but for the F555W filter.}
\label{fig:555binaries_1}
\end{figure*}

\begin{figure*}[htb]
\gridline{\fig{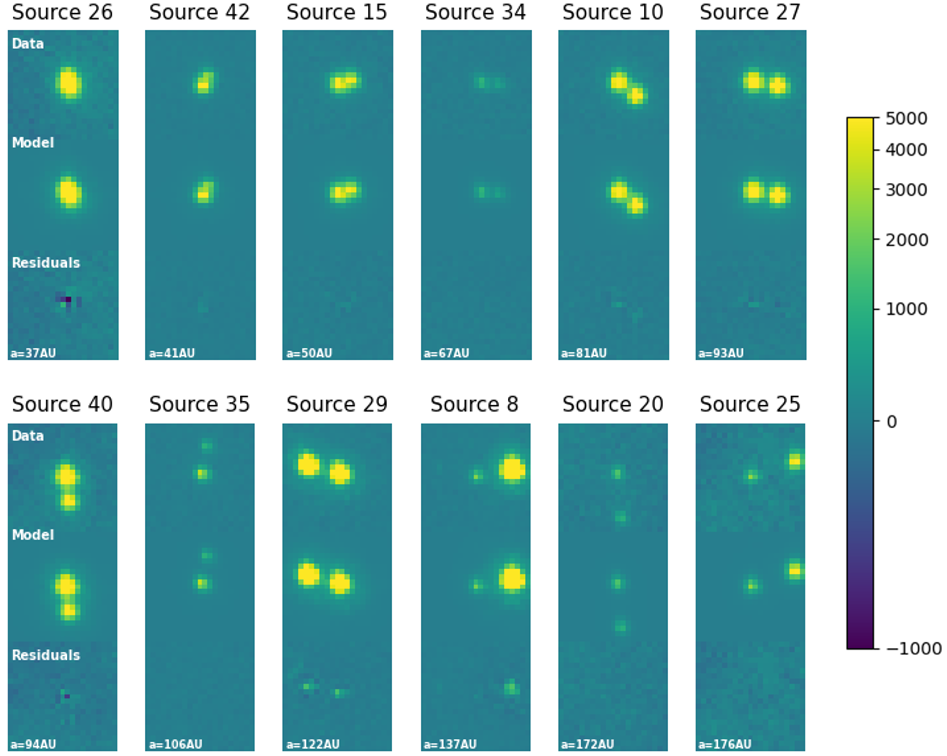}{\textwidth}{}}
\caption{Continuation of Fig. \ref{fig:555binaries_1}.}
\label{fig:555binaries_2}
\end{figure*}

\begin{figure*}[htb]
\gridline{\fig{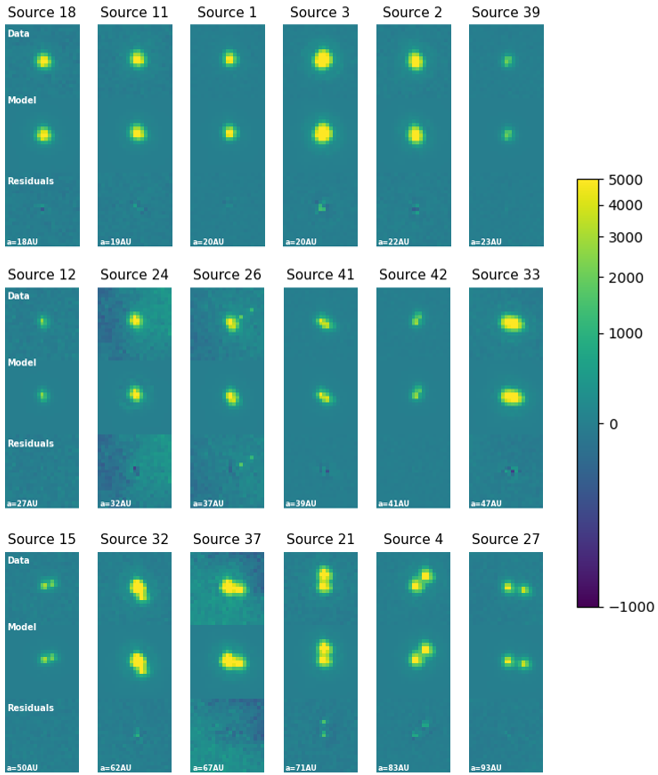}{\textwidth}{}}
\caption{Same as Fig. \ref{fig:435binaries_1}  but for the F658N filter.}
\label{fig:658binaries_1}
\end{figure*}

\begin{figure*}[htb]
\gridline{\fig{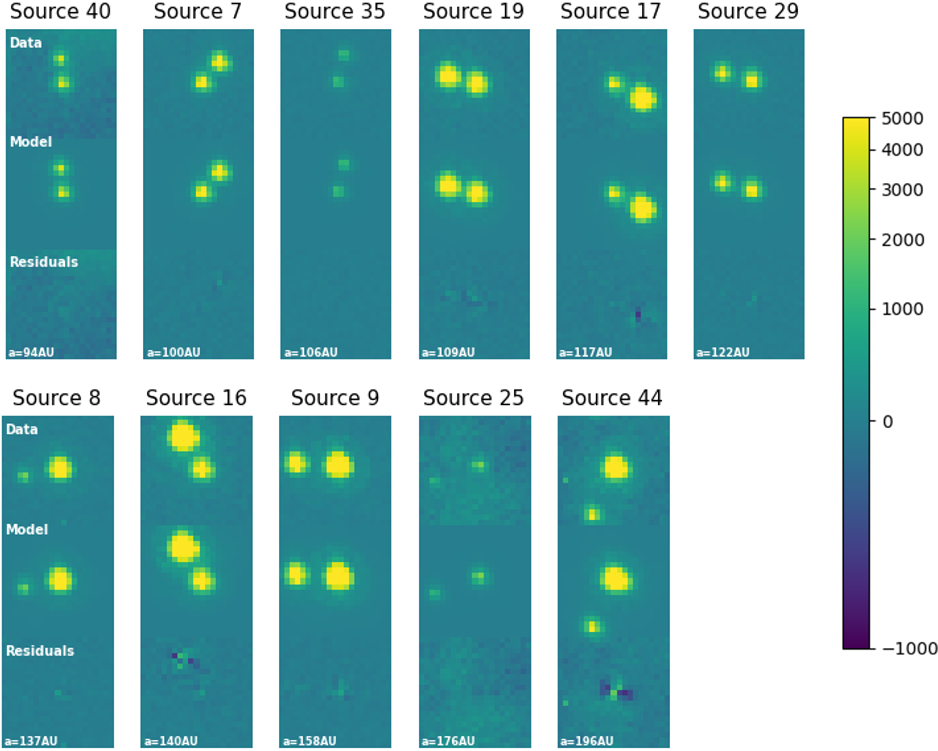}{\textwidth}{}}
\caption{Continuation of Fig. \ref{fig:658binaries_1}.}
\label{fig:658binaries_2}
\end{figure*}

\begin{deluxetable*}{cccccccc}
\tablenum{3}
\tablecaption{All sources within the sample, listed with their highest signal-to-noise (S/N) within all images of the HST Treasury Program data.}
\tablewidth{0pt}
\tablehead{
\colhead{2MASSID} & \colhead{S/N} & \colhead{2MASSID} & \colhead{S/N} &
\colhead{2MASSID} & \colhead{S/N} & \colhead{2MASSID} & \colhead{S/N}
}
\startdata
    J05341202-0524196 & 164 & J05342080-0523291 & 302 & J05342650-0523239 & 370 & J05342698-0518033 & 134\\
    J05342753-0528284 & 517 & J05342926-0523509 & 54 & J05342949-0513551 & 506 & J05342954-0523437 & 435\\
    J05343168-0528269 & 587 & J05343202-0527426 & 240 & J05343955-0527174 & 133 & J05344083-0528095 & 350\\
    J05344184-0534299 & 239 & J05344286-0525163 & 311 & J05344441-0526061 & 341 & J05344656-0523256 & 299\\
    J05344677-0526048 & 304 & J05344679-0521291 & 376 & J05344791-0535438 & 315 & J05344828-0532351 & 171\\
    J05344877-0519073 & 343 & J05344878-0517464 & 389 & J05344896-0528168 & 321 & J05344907-0526266 & 451\\
    J05345009-0517121 & 100 & J05345085-0529250 & 513 & J05345099-0517565 & 195 & J05345120-0516549 & 554\\
    J05345201-0524187 & 402 & J05345233-0530080 & 499 & J05345259-0515366 & 215 & J05345265-0529452 & 670\\
    J05345275-0527545 & 510 & J05345359-0526371 & 501 & J05345418-0528543 & 580 & J05345483-0525125 & 250\\
    J05345555-0536061 & 245 & J05345560-0529375 & 275 & J05345583-0519454 & 65 & J05345675-0526372 & 209\\
    J05345683-0521363 & 414 & J05345693-0522062 & 195 & J05345701-0523000 & 256 & J05345714-0533294 & 437\\
    J05345737-0514334 & 214 & J05345792-0529460 & 195 & J05345802-0517376 & 327 & J05345826-0538257 & 41\\
    J05345827-0525332 & 196 & J05345837-0521166 & 183 & J05345853-0532498 & 231 & J05345879-0521176 & 307\\
    J05345893-0513455 & 127 & J05345918-0523078 & 247 & J05345931-0523326 & 182 & J05350024-0518508 & 352\\
    J05350101-0524103 & 401 & J05350116-0529551 & 715 & J05350121-0521444 & 38 & J05350129-0520168 & 375\\
    J05350133-0520221 & 267 & J05350148-0528207 & 474 & J05350160-0524101 & 307 & J05350161-0533380 & 190\\
    J05350201-0518341 & 126 & J05350207-0518226 & 148 & J05350218-0529098 & 487 & J05350243-0520465 & 605\\
    J05350270-0532249 & 110 & J05350274-0519444 & 301 & J05350284-0522082 & 106 & J05350309-0522378 & 255\\
    J05350315-0518299 & 380 & J05350322-0517532 & 147 & J05350332-0516227 & 606 & J05350370-0522457 & 25\\
    J05350396-0518597 & 331 & J05350416-0520156 & 55 & J05350434-0538311 & 183 & J05350437-0523138 & 689\\
    J05350450-0526044 & 323 & J05350461-0524424 & 146 & J05350476-0517421 & 373 & J05350481-0522387 & 326\\
    J05350487-0520574 & 80 & J05350495-0521092 & 587 & J05350506-0536438 & 638 & J05350513-0520244 & 453\\
    J05350537-0524105 & 249 & J05350540-0524150 & 128 & J05350560-0518248 & 161 & J05350561-0529223 & 531\\
    J05350571-0523540 & 283 & J05350584-0527016 & 347 & J05350588-0527090 & 138 & J05350609-0514249 & 180\\
    J05350615-0519556 & 322 & J05350617-0522124 & 37 & J05350627-0522027 & 213 & J05350642-0527048 & 264\\
    J05350727-0522266 & 235 & J05350732-0538409 & 48 & J05350739-0525481 & 238 & J05350744-0526401 & 379\\
    J05350768-0536587 & 303 & J05350773-0521014 & 311 & J05350784-0529174 & 670 & J05350803-0536140 & 346\\
    J05350822-0524032 & 42 & J05350829-0524348 & 161 & J05350834-0527569 & 325 & J05350838-0528293 & 498\\
    J05350859-0526194 & 106 & J05350870-0529016 & 465 & J05350873-0522566 & 60 & J05350920-0530585 & 99\\
    J05350959-0527599 & 426 & J05350976-0521282 & 136 & J05350985-0519339 & 163 & J05351015-0527574 & 146\\
    J05351021-0523215 & 153 & J05351029-0519563 & 176 & J05351031-0521130 & 223 & J05351041-0519523 & 92\\
    J05351047-0526003 & 72 & J05351050-0522455 & 282 & J05351057-0533136 & 295 & J05351073-0526280 & 244\\
    J05351083-0525569 & 80 & J05351088-0528007 & 370 & J05351121-0517209 & 498 & J05351163-0522515 & 153\\
\enddata
\end{deluxetable*}

\begin{deluxetable*}{cccccccc}
\tablenum{4}
\tablecaption{Continuation of Table 3}

\tablewidth{0pt}
\tablehead{
\colhead{2MASSID} & \colhead{S/N} & \colhead{2MASSID} & \colhead{S/N} &
\colhead{2MASSID} & \colhead{S/N} & \colhead{2MASSID} & \colhead{S/N}
}
\startdata
    J05351165-0524213 & 169 & J05351165-0531011 & 340 & J05351178-0521555 & 159 & J05351185-0517259 & 119\\
    J05351188-0521032 & 84 & J05351197-0522541 & 156 & J05351216-0530201 & 503 & J05351227-0520452 & 166\\
    J05351256-0516332 & 186 & J05351269-0519353 & 82 & J05351270-0527106 & 265 & J05351277-0520349 & 91\\
    J05351282-0539077 & 566 & J05351294-0528498 & 243 & J05351301-0519041 & 350 & J05351303-0534035 & 184\\
    J05351304-0520302 & 256 & J05351305-0521532 & 41 & J05351324-0527541 & 480 & J05351330-0520189 & 143\\
    J05351336-0522261 & 150 & J05351343-0521073 & 295 & J05351345-0517103 & 454 & J05351348-0530481 & 514\\
    J05351351-0517173 & 418 & J05351352-0527286 & 391 & J05351356-0527573 & 198 & J05351357-0535080 & 324\\
    J05351365-0528462 & 314 & J05351375-0534548 & 325 & J05351379-0519254 & 210 & J05351382-0527368 & 269\\
    J05351389-0518531 & 99 & J05351397-0521233 & 69 & J05351421-0520042 & 183 & J05351444-0533190 & 509\\
    J05351445-0517254 & 664 & J05351465-0523018 & 138 & J05351475-0534167 & 193 & J05351491-0536391 & 690\\
    J05351534-0519021 & 400 & J05351545-0517383 & 368 & J05351547-0527227 & 366 & J05351548-0535118 & 368\\
    J05351559-0534466 & 236 & J05351567-0517472 & 82 & J05351569-0528155 & 120 & J05351571-0526283 & 279\\
    J05351587-0522328 & 75 & J05351596-0516575 & 269 & J05351609-0524112 & 110 & J05351624-0528337 & 25\\
    J05351627-0532021 & 260 & J05351632-0515380 & 357 & J05351661-0519357 & 185 & J05351676-0517167 & 75\\
    J05351689-0517029 & 36 & J05351694-0525469 & 338 & J05351700-0515443 & 346 & J05351712-0524585 & 132\\
    J05351736-0520149 & 204 & J05351743-0530253 & 528 & J05351751-0517401 & 212 & J05351778-0523155 & 193\\
    J05351789-0518352 & 211 & J05351794-0525061 & 58 & J05351794-0525338 & 213 & J05351795-0535157 & 560\\
    J05351797-0516451 & 169 & J05351797-0526506 & 88 & J05351809-0515461 & 417 & J05351820-0516340 & 284\\
    J05351820-0524302 & 140 & J05351826-0529538 & 382 & J05351851-0520427 & 194 & J05351858-0526248 & 143\\
    J05351873-0518024 & 127 & J05351884-0522229 & 449 & J05351894-0520522 & 318 & J05351921-0531030 & 69\\
    J05351979-0530376 & 416 & J05351983-0515089 & 374 & J05351986-0530321 & 349 & J05351986-0531038 & 162\\
    J05352002-0529119 & 72 & J05352017-0523085 & 215 & J05352032-0536394 & 179 & J05352041-0517144 & 285\\
    J05352054-0524208 & 95 & J05352067-0523531 & 147 & J05352082-0521216 & 122 & J05352099-0516375 & 509\\
    J05352103-0522250 & 77 & J05352104-0523490 & 38 & J05352115-0518213 & 173 & J05352162-0526576 & 129\\
    J05352165-0517173 & 504 & J05352172-0526443 & 255 & J05352184-0522082 & 154 & J05352190-0515011 & 166\\
    J05352192-0528273 & 171 & J05352194-0517043 & 363 & J05352206-0528152 & 300 & J05352228-0531168 & 412\\
    J05352246-0525451 & 160 & J05352266-0515085 & 96 & J05352268-0516140 & 255 & J05352279-0531372 & 643\\
    J05352296-0522415 & 227 & J05352303-0529414 & 435 & J05352312-0513435 & 33 & J05352317-0522283 & 572\\
    J05352321-0521357 & 129 & J05352331-0528100 & 183 & J05352332-0521254 & 311 & J05352349-0520016 & 100\\
    J05352376-0518398 & 515 & J05352396-0519076 & 366 & J05352431-0528441 & 351 & J05352433-0526003 & 181\\
    J05352445-0524010 & 114 & J05352463-0519096 & 493 & J05352465-0522425 & 37 & J05352488-0525101 & 218\\
    J05352512-0522252 & 102 & J05352522-0529516 & 419 & J05352523-0533210 & 259 & J05352534-0525295 & 203\\
    J05352537-0524114 & 165 & J05352543-0521515 & 241 & J05352547-0521349 & 358 & J05352547-0534028 & 360\\
    J05352568-0530381 & 436 & J05352571-0523094 & 377 & J05352605-0521210 & 192 & J05352615-0522570 & 50\\
    J05352616-0520060 & 401 & J05352618-0525203 & 99 & J05352630-0527439 & 258 & J05352640-0516124 & 65\\
\enddata
\end{deluxetable*}

\end{document}